\documentclass[usenatbib]{mn2e}
\usepackage{rotating}

\begin{document}

\newcommand{\GHz}{\rm GHz}
\newcommand{\MHz}{\rm MHz}
\newcommand{\h}{^{\rm h}}
\newcommand{\m}{^{\rm m}}
\newcommand{\s}{^{\rm s}}
\newcommand{\pSqa}{{\rm arcsec}^{-2}}
\newcommand{\lta}{\raisebox{-0.4ex}{$\,\stackrel{<}{\scriptstyle \sim}\,$}} 
\newcommand{\gta}{\raisebox{-0.4ex}{$\,\stackrel{>}{\scriptstyle \sim}\,$}}

\title[Faint 1.4~GHz sources in 2dFGRS]
{Faint 1.4~GHz sources in the 2dF Galaxy Redshift Survey}

\author[Chan, Cram, Sadler, Killeen, Jackson, Mobasher \& Ekers]
{B.H.P. Chan$^{1,\star}$, L.E. Cram$^1$, E.M. Sadler$^1$, N.E.B. Killeen$^2$, 
\newauthor C.A. Jackson$^{2,3}$, B. Mobasher$^4$, R.D. Ekers$^2$.\\ 
$^1$School of Physics, A28, University of Sydney, NSW 2006, Australia\\ 
$^2$Australia Telescope National Facility, CSIRO, Epping, NSW 1710, Australia\\ 
$^3$Research School of Astronomy \& Astrophysics, The Australian National 
    University, Mount Stromlo Observatory, Canberra, Australia\\ 
$^4$Space Telescope Science Institute, 3700 San Martin Drive, Baltimore, MD 
    21218, USA\\
$^\star$bchan@physics.usyd.edu.au\\
} 
\maketitle

\begin{abstract}
The Australia Telescope Compact Array (ATCA) has been used to survey 
at 1.4~GHz, a small region ($<$ 3 sq deg) overlapping with the 2dF 
Galaxy Redshift Survey \citep{Colless01}. We surveyed with a varying 
radio sensitivity, ranging from 1~mJy -- 20~$\mu$Jy ($1\sigma$). There 
are 365 2dFGRS sources with z $>$ 0.001 lying within the surveyed 
region, of which 316 have reliable spectral classification. Following 
\citet{Sadler02}, we visually classified 176 as AGN or early-type 
galaxies, and 140 as star-forming galaxies. We derived radio 
flux density measurement or upperlimits for each of the 365 2dFGRS 
sources. The fraction of radio 
detected 2dFGRS star-forming galaxies increases from $\sim$ 50\% at 
$\sim$ 0.7~mJy up to $\sim$ 60\% at $\sim$ 0.2~mJy. The mean redshift 
for the fraction of radio detected star-forming galaxies increases with 
increasing radio detection sensitivity, while the mean redshift is 
fairly constant for the AGN/early-type fraction. We found very 
similar radio detection rates of 2dFGRS galaxies for both the 
AGN/early-type and star-forming components. The radio detection rate 
increases approximately linearly with respect to the rate of increase 
in radio detection sensitivity. We derived the radio luminosity function 
for our sample and it was found to be consistent with that of \citet{Sadler02}. 
We have also compared the total flux densities of NVSS sources common 
to our survey, and we discuss strategies for a large-scale radio survey 
of the 2dFGRS sample.
\end{abstract}

\begin{keywords}
surveys -- galaxies: statistics -- galaxies: evolution -- 
galaxies: distances and redshifts -- 
galaxies: fundamental parameters --
radio continuum: galaxies
\end{keywords}

\section{Introduction}
The decimetric synchrotron emission from galaxies is believed to arise
from relativistic electrons accelerated by shock waves either in the
interstellar medium, or in structures associated with an active galactic
nucleus (AGN). In the former case, there is evidence that the
decimetric luminosity is proportional to the rate of massive star
formation, the link arising because supernovas are perhaps the
dominant source of accelerating shock waves \citep{Biermann76,Condon92}. 
In the later case, the decimetric luminosity of an AGN may 
be related to the physical properties of the underlying black hole, 
and the rate and mode of mass accretion by that black hole 
\citep{Franceschini98,Laor02,Ho02}.

Observations of the total decimetric luminosity of a galaxy thus allow
an estimate of the extinction-free star-formation rate in galaxies
where AGN emission can be considered negligible. Conversely, where the 
star formation 
rate is relatively low they provide information on the character of
activity excited by a nuclear black hole. A few galaxies emit
significant decimetric from combined nuclear star-formation and AGN
activity \citep{Hill01,Veilleux01,Nagar02}.

The knowledge provided by radio flux density measurements is greatly
enhanced when optical spectroscopic data are also available. Optical
spectra reveal the galactic redshift, allowing the calculation of
luminosities and hence estimates of the cosmic evolution of the rates
of star-formation and black hole mass accretion. They also reveal the
age and other properties of the most optically luminous components of
the galactic stellar population and, in some cases, the state of
excitement of a substantial fraction of the interstellar medium of the
galaxy. Investigations of large galaxy samples having combined optical
and radio data offer opportunities to address several outstanding
astrophysical problems.

The 2dF Galaxy Redshift Survey \citep[2dFGRS,][]{Colless01} has measured
the optical spectra of over 220,000 galaxies photometically selected
to have an extinction-corrected magnitude brighter than $b_J$=19.45.
The survey comprises strips at the equator, and near the South
Galactic Pole (SGP). The SGP strip densely covers $\sim 80^\circ
\times ~ 12^\circ$ ($12\h40\m < \alpha < 03\h40\m$, $-36.5^\circ <
\delta < -24.5^\circ$), along with 99 randomly placed $2^\circ$ (diameter) fields.

\citet{Sadler02} studied the relatively bright radio sources among the
2dFRGS objects by cross-matching the 1.4~GHz NRAO VLA Sky Survey
\citep[NVSS,][]{Condon98} with the first $\sim$ 20\% of the 2dFGRS.
Approximately 1.5\% of the 2dFGRS galaxies ($\sim$ 5\% NVSS sources) 
have radio counterparts at the NVSS radio flux density detection limit 
of 2.8~mJy. Presumably, the majority of unmatched (95\%) NVSS 1.4~GHz 
sources are AGNs whose optical counterparts are too faint to be detected 
at $b_J$=19.45 \citep[c.f.][]{Condon88}. Some 60\% of the identifications 
are AGNs (radio galaxies and some Seyferts) and the remainder 40\% 
comprised of star-forming objects. \citet{Sadler02} note that complete 
cross-matching of the 2dFGRS and NVSS will provide approximately 4,000 
radio source spectra, a sample large enough to measure radio galaxy 
evolution to
$z=0.35$ and to locate the most luminous star-forming galaxies to
$z=0.2$. \citet{Condon02} have reported evidence at the level of two
standard deviations for evolution in the rate of star formation density 
between the mean redshifts, $\langle z \rangle \sim$ 0.02 and 0.06, of 
their NVSS/UGC sample and the NVSS/2dFGRS sample of \citet{Sadler02}, 
respectively.

There is considerable interest in seeking radio identifications of
2dFRGS objects below the NVSS flux density limit. Spectroscopy of the
optical counterparts of radio sources found in sub-mJy surveys 
\citep[e.g.][]{Benn93,Georgakakis99,Gruppioni99,Prandoni01}
suggests that the faint radio population will be a mixture of star-forming
galaxies and AGNs, as it is for the mJy radio identifications.
Interestingly, 1.4~GHz radio source counts rise sharply towards the
Euclidean rate near 1~mJy \citep[e.g.][]{Condon88}, a phenomenon
ascribed to the rising proportion of star-forming galaxies in the
sub-mJy population \citep{Windhorst84,Condon88,Benn93,Hopkins98}. 
In view of the tight correlation between radio and far infra-red (FIR) 
luminosity among the star-forming population, these galaxies likely 
correspond to the population of lower star formation rate ``IRAS galaxies'' 
seen at apparently brighter magnitudes. Studies of the sub-mJy 
counterparts of the 2dFGRS objects will thus encompass star-forming 
objects down to $\sim 10^{22.5}$~WHz$^{-1}$ out to redshifts of $z \sim 0.1$,
where evolution appears to have occurred, as well as AGNs of moderate radio 
luminosity.

The SGP part of 2dFGRS covers over 900 square degrees and would
require a very large allocation of radio telescope time to undertake a
co-extensive sub-mJy survey. Acknowledging this, we have undertaken a
pilot study of a selected region of the 2dFGRS (equivalent area of one 
2dFRGS-field), employing the Australian Telescope Compact Array (ATCA) 
at 1.4~GHz to explore probable outcomes and optimal observing strategies 
for a more extensive survey.

Section $\S2$ describes the survey field selection, observations and
data reduction and $\S3$ the procedures used to derive radio flux
measurements and radio upper limits of the 2dFGRS objects. $\S4$ presents 
the radio luminosity function derived from our sample and comparison with 
that of \citet{Sadler02}. In $\S5$ we explore the radio detection rate of 
2dFGRS sources as a function of survey depth and in $\S6$ we consider the 
optimisation of a full-scale sub-mJy survey.

Throughout the paper we have used $\Omega_{\rm M} = 1$, $H_o =
70$kms$^{-1}$Mpc$^{-1}$, and adopted for all sources, a radio spectral 
index $\alpha = -0.7$ ($S\propto \nu^\alpha$).

\section{Survey strategy, field selection} 

One aim of this pilot survey is to investigate a survey strategy which
yields optimal benefit from the ATCA observing time that would be
invested in a deep, large-scale radio follow-up of the 2dFGRS. The
basic trade-off lies between the steeply rising surface density of
sources at fainter survey limits, versus the quadratic dependence of
survey sensitivity on integration time. This implies a strategy in
which a small area is surveyed very deeply, and a progressively larger
area is surveyed to a progressively poorer sensitivity. Given the
breadth of the total radio luminosity functions for both star-forming
and AGN galaxies, and the weak dependence of the bi-variate radio
luminosity function on optical luminosity, it would be hard to argue
that any flux density range is to be preferred over any other,
suggesting a strategy in which equal numbers of sources are detected
to a progressively changing flux density limit.

Table~1 presents the results of calculations designed
to illustrate this strategy. The Table uses the published sensitivity
of the ATCA at 1.4~GHz using two 128~MHz IF channels, namely that a
point source of $S_{lim}$~mJy is detected at the field centre with
$5\sigma$ sensitivity in a time of $(2.24/S_{lim})^2$ minutes,
provided that full $(u,v)$ coverage is obtained (ATCA on-line
sensitivity calculator\footnote{http://www.atnf.csiro.au/observers/docs/at\_sens/}). 
The Table relies on integral source densities read from Fig.~2(b) in 
\citet{Becker95}. According to the Table, a single pointing observed for 5,000~m 
(83~h) would reveal just over 3,000 sources per square degree, brighter than 
30~$\mu$Jy, and 
approximately the same number of sources would be detected to a 
$5\sigma$ limit of 1~mJy if an area of 34 pointings was surveyed for 
a total of 170~m (3~h). These values are ideal estimate, excluding observation 
overheads. In practice, there are several other factors that must be considered 
in designing the survey. The limits given in the table refer to the most 
sensitive part of the primary beam, and can be approached to within a factor 
of about $\sqrt{2}$ over a large area only if mosaicing techniques are
exploited. Observations must be staged to provide adequate $(u,v)$
coverage, and bright radio sources must be avoided to achieve high
dynamic range.

\begin{table} 
\begin{center} 
\caption{Observing strategy.}
\label{tab:design} \vspace{1em} 

\begin{tabular}{lrrrrr} 
\hline 
(1) & (2) & (3) & (4) & (5) & (6) \\
Sensitivity & $1\sigma$ & time  & $N>S$        & Relative & Relative \\
($\mu$Jy)   & ($\mu$Jy) & (min) & (deg$^{-2}$) & area     & time     \\ \hline 
$>$1000 & 200 & 5 & 90 & 34 & 1 \\ 
$>$316  &  63 & 50 & 275 & 11 & 3 \\ 
$>$100  &  20 & 500 & 900 & 3.4 & 10 \\ 
$>$ 31  &   6 & 5000 & 3000& 1 & 30  \\ 
\hline

\end{tabular}\\ 
\end{center} 
(1) Point source $5\sigma$ sensitivity.\\
(2) $1\sigma$ sensitivity.\\
(3) Time to reach $1\sigma$.\\
(4) Surface density of 1.4~GHz sources brighter than $5\sigma$ limit.\\
(5) Relative areas for equal source numbers.\\
(6) Relative observing time for specified sensitivity and source numbers.\\

\end{table}
 
Awarded $8 \times 12$ hours of observation time for the pilot survey,
we chose a mosaic of 16 pointings in square tessellation with a pitch
of 17 arcmin.  Embedded within the 16 pointing mosaic are sub-mosaics
of 8, 4 and 2 pointings. Fig.~1 illustrates our
mosaic pattern, where denser filling indicates a longer integration
time at that pointing. This strategy is not a precise match to the
data in Table~1, but it suffices to illustrate and
explore it.

There are several requirements to consider in selecting the field.
Firstly, it should lie away from strong sources ($S \geq$ 100~mJy), and
be devoid of sources $\geq$ 50~mJy in the deepest region in order to
achieve the desired sensitivity. The NVSS has been used to identify
regions relatively free of strong sources. Secondly, it must lie
within 2dFGRS and, for testing purposes, have at least 50\% redshift
completeness in the 100k 2dFGRS release \citep{Colless01}. Finally, 
we note that the 2dFGRS SGP strip passes through the zenith of the
ATCA (latitude $-30^\circ18^\prime52\arcsec$). Since mosaicing near
the zenith results in large slew times for the alt-az mount, we choose
to observe as far away from $\delta \sim -30^\circ$ as possible.

Using these criteria we place the survey centre at $\alpha =
23\h37\m35\s, \delta = -27^\circ52^\prime45\arcsec$ (J2000), 
coincident with the 2dF fields SGP192--SGP194 \& SGP270--SGP272. 
Fig.~2 displays two adjoined NVSS mosaics (I2344M28 \&
I2328M28) showing the neighborhood of the target area.  Annotated on
this image are (i) large circles scaled to represent the 16 pointings
of the 33 arcmin FWHM of the 1.4~GHz ATCA primary beam, (ii) small
circles representing NVSS sources with 50~mJy $\leq S <$ 100~mJy,
(iii) squares representing NVSS sources with 100~mJy $\leq S <$
150~mJy and (iv) numbered squares representing NVSS sources with $S
\geq$ 150~mJy.  All NVSS sources with $S \geq$ 50~mJy lying within a
2$^\circ$ radius of the ATCA mosaic centre are marked in the image.

\begin{figure}
\begin{center}
\noindent\includegraphics[width=.2\textwidth]{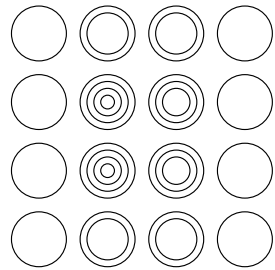}
\caption{Illustration of mosaic pattern and survey depth, where each 
concentric circle groups represents individual pointing centres. In 
accordance with the denotation in $\S2.1$, from lesser to denser 
filling circles indicating increase of total integration times 
corresponding to fields denoted A-fields up to D-fields.} 
\label{fig:pointings}
\end{center}
\end{figure}

\begin{figure}
\begin{center}
\noindent\includegraphics[width=.45\textwidth]{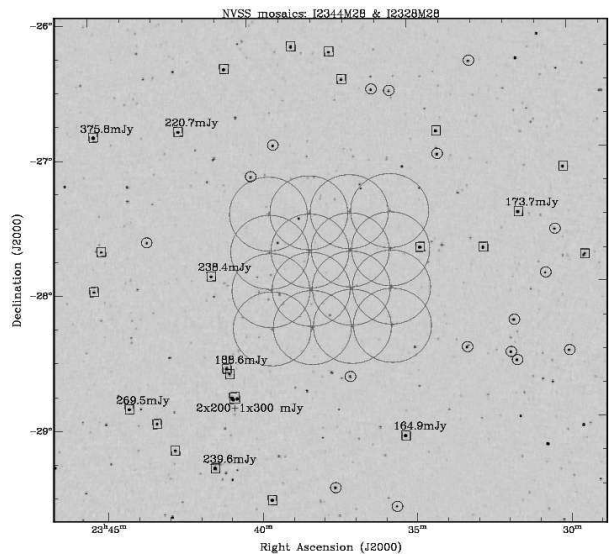}
\caption{NVSS image. Large circles shows the pointings (FWHM) surveyed with
the ATCA. We have marked all NVSS sources within a 2$^\circ$ radius of 
the ATCA mosaic centre having 50~mJy $\leq S <$ 100~mJy with small 
circles, and 100~mJy $\leq S <$ 150~mJy sources with small squares. 
Sources with $S \geq$ 150~mJy are explicitly indicated. The greyscale 
spans -10 to 40~mJy beam$^{-1}$.}
\label{fig:NVSS}
\end{center}
\end{figure}

\subsection{Observations and data reduction}

Observations were performed over 8 days at the end of May 2001
with the ATCA in
the 6F configuration. We used 2 $\times$ 128~MHz IFs centred at
1344~MHz and 1432~MHz. Denoting the 16, 8, 4 and 2 pointing
sub-mosaics as fields A, B, C and D, respectively,
Table~2 list the average total integration time for
each of these fields. To simplify observing programs, A-fields were
observed only during a single 1 $\times$ 12~h session, B-fields
only in 2 $\times$ 12~h sessions, C-fields in 4 $\times$ 12~h sessions
and D-fields observed in all 8 sessions.

To ensure good image quality, we need adequate $(u,v)$ coverage. The
6~km array rotates to an independent $(u,v)$ cell in about 80~s and
ideally we would take two samples within this time.  With 10~s
integrations and the relevant slew times, this can be achieved only
when mosaicing at most 4 pointings, especially since slew times 
increase for fields that transit near the zenith. Since we have
16 pointings in the A-fields, all obtained on a single day, they are
under sampled. However the other fields are properly sampled since we
have multiple days of observations to fill the $(u,v)$ plane.

For primary calibration we used the ATCA standard PKS
B1934-638, adopting $S_{1384~\MHz}$ = 14.9~Jy
\citep{Baars77}. For secondary calibration we have used
B2331-240, adopting $S_{1.5~\GHz}$ = 1~Jy \citep{Taylor02}.
The mosaic cadence is broken for secondary calibration every 40-50~m,
scanning the calibrator for 2~m.

\begin{table}
\begin{center}
\caption{Observation summary, with field-labels defined in $\S2.1$,
and illustrated in Fig.~1.}
\label{tab:obssum}
\vspace{1em}
\begin{tabular}{lrrr} \hline
Field & $\tau^\star$ (hr) & \% flagged & $\sigma_T^\dagger$
($\mu$Jy/beam) \\ \hline A & 0.45 & 8 & 105 (0.9K) \\ B & 1.62 & 5 &
55 (0.5K) \\ C & 6.32 & 5 & 28 (0.2K) \\ D & 25.30 & 5 & 14 (0.1K) \\
\hline
\end{tabular}\\
\end{center}
$^\star$ Averaged integration time.\\ $\dagger$ Theoretical rms noise
level and brightness temperature sensitivity. 
\end{table}

\subsubsection{Calibration and flagging}
The 8 days of observations and each of the two IFs were reduced separately
following the standard {\sc MIRIAD} data reduction procedures:

\begin{enumerate} 

\item \textit{Primary flux calibration}: The primary
calibrator $(u,v)$ data set were flagged to remove bad $(u,v)$ samples and
the gain, bandpass ({\sc mfcal}) and leakage ({\sc gpcal}) calibration
tables were calculated using a 10~s solution interval.  

\item \textit{Secondary phase calibration}: As the target field was
near the zenith there is limited coverage of parallactic angle for the
secondary calibrator. Thus, the bandpass and leakage calibration
tables from the primary calibrator are used ({\sc gpcopy}).  The
secondary calibrator $(u,v)$ data set were then flagged and the gain
calibration determined ({\sc gpcal}) using a 10~s solution interval.

\item \textit{Program source}: Calibration tables are copied to each
of the program sources (pointings) from the secondary calibrator ({\sc
gpcopy}) and each of the $(u,v)$ data sets flagged.

\end{enumerate} 

For our data flagging, we used the task {\sc tvclip}
developed by \citet{Prandoni00} (see their $\S4.1$). This allows the
user to recursively flag a $(u,v)$ data set by specifying (i) a specific
quantity from which a running median of the data sample is determined
(e.g.  amplitude of a polarisation) and (ii) a clip level, so that any
$(u,v)$ data point exceeding the running median by more than the factor
specified by the clip level will be flagged. We have found that
deleting $(u,v)$ points having an amplitude greater than 10 times the
running median removes typical interference.  Where lower level
interference persists we repeat the flagging procedure and remove all
samples greater than 6 times the running median.
Table~2 lists the averaged (all days and both IFs)
percentage of data flagged for the field types defined in $\S2.1$. The
overall total amount of data flagged is 4.5\%.

\subsubsection{Imaging and self-calibration}

Once the data are calibrated and interference is excised, we proceed
to masaicing.  We have adopted a joint deconvolution approach with a
Steer-Dewdney-Ito (SDI) CLEAN algorithm \citep{Steer84} using the task
{\sc mossdi2}.  This task is a version of {\sc mossdi} modified by
\citet{Regan01} so that the CLEAN-cutoff level is expressed in terms
of the locally determined image RMS instead of a fixed value in Jy
beam$^{-1}$.  This feature was very important in view of the large
variation in the total integration time per pointing across the
mosaiced region.  We followed the basic {\sc MIRIAD}
joint-deconvolution procedures:

\begin{enumerate}

\item Apply {\sc invert} to the $(u,v)$ data of all pointings, both IFs
(invoking options \textit{mfs} to deal with bandwidth smearing) and
all 8 days, with uniform weighting, 2 arcsec cell size, to generate a
dirty mosaic and dirty beam cube (one dirty beam image per pointing).

\item Use {\sc mossdi2} to generate a joint-deconvolution
CLEAN-component model of the dirty mosaic, CLEANing down to a
specified CLEAN-cutoff level.

\item Use {\sc restor} to subtract the CLEAN component model,
convolved with the appropriate dirty beam, from the dirty mosaic and
then restore with an elliptical Gaussian profile fitted from the dirty
beam.

\item Lastly, use {\sc mossen} to generate a gain image which then
allows primary beam correction over the entire mosaic.

\end{enumerate}

There are two artefacts evident in the mosaic. Firstly, weak positive
and negative rings appear around strong sources, presumably reflecting
amplitude calibration errors. The effect is largest for bright sources
lying on the flank of the primary beam and is probably caused by
pointing errors. There are also radial artifacts appearing as
symmetric and asymmetric radial structures emerging from strong
sources.  These can be removed by phase self calibration. Comparison 
of our mosaics before and after the self calibration procedures, we 
found no spurious sources above 3$\sigma$ level, and source flux 
densities varies less than $\sim$ 5\%, much smaller than the expected 
fitting uncertainty. The following summarises our imaging steps: 

\begin{enumerate}

\item Without restricting the CLEAN region deconvolve
with {\sc mossdi2}, CLEANing down to a CLEAN-cutoff level of
10$\sigma$ and generate a preliminary mosaic. 

\item Use automatic source fitting of the preliminary mosaic to
generate a list of CLEAN boxes over all fitted sources with peak flux
density greater than 10 times the local image noise rms.

\item Generate a CLEAN component model for self-calibration using {\sc
mossdi2}, CLEANing over the boxes defined in (ii).  Invoke the option
\textit{positive} to constrain the CLEAN components to be positive
valued.

\item With the CLEAN components defined in (ii) investigate image
quality improvement using various self-calibration solution intervals.
Radial artifacts were completely removed for solution intervals lying
between 10-30~s.  We adopted 30~s to enhance the signal-to-noise ratio
per self-calibration solution interval.

\item With {\sc selfcal} and a 30~s solution interval apply phase
self-calibration to each pointing, each IF, and each day separately.

\item Using self-calibrated $(u,v)$ data, deconvolve with {\sc mossdi2}
CLEANing down to 5$\sigma$ over CLEAN boxes defined in (ii). Use the
new mosaic to define a new set of CLEAN boxes for all fitted sources
with peak flux density greater than 5 times the local image noise
rms. Finally generate a deep-CLEANed mosaic with {\sc mossdi2},
CLEANing down to 2$\sigma$ over the new CLEAN boxes.

\end{enumerate}

Displayed in Fig.~3 is the deep-CLEANed mosaic with an
enlarged central portion. The contours represent $1\sigma$ local image
noise levels calculated in the manner described below. The 0.2~mJy
noise contour in Fig.~3 corresponds closely to the
outline of the 16-pointing mosaic region illustrated in
Fig.~2. The synthesised beam size of the deep-CLEANed
mosaic is $15.5\arcsec \times 8.0\arcsec$ at position angle
$1.20^\circ$.

\begin{figure}
\begin{center}
\noindent\includegraphics[width=.45\textwidth]{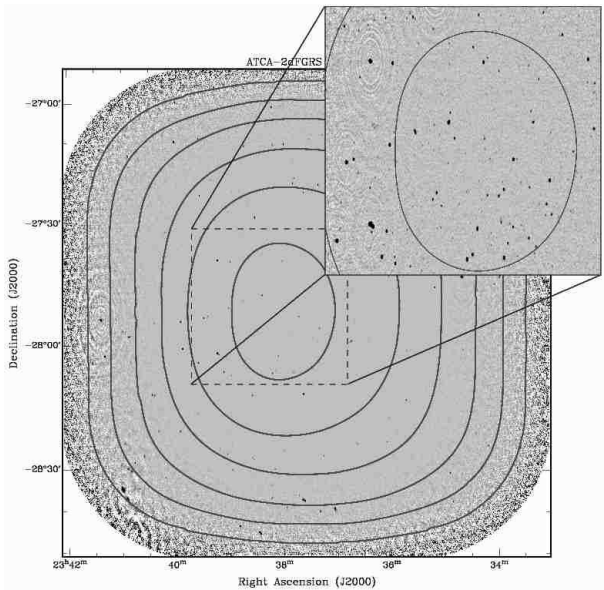}
\caption{ATCA mosaic.  The contours are drawn at the 1$\sigma$ noise
level for 0.025, 0.05, 0.1, 0.2, 0.4 and 1~mJy.  The greyscale is set
to -1.5 to 4.5~mJy beam$^{-1}$ for full the mosaic and -0.15 to 
0.45~mJy beam$^{-1}$ for the inset.}
\label{fig:ATCA}
\end{center}
\end{figure}

\subsection{Local image noise estimate}
It is important to estimate the local RMS noise for the image used in the
source-fitting phase of our study. We could estimate the noise using
{\sc restor} to form a residual image of the deep-CLEANed mosaic, but
the estimate would exhibit irregular spatial variations.  Instead we
have generated a theoretical sensitivity image using {\sc mossen}, and
checked the predictions using noise estimates based on 1000 randomly
selected patches each $\sim 3 \times$ the synthesised beam area. 
The tight correlation between the prediction and the measured
values shown in Fig.~4 is fit by $\sigma =
10^{-0.06} \langle n \rangle^{0.97}$ with $\sim$ 20\% uncertainty. This equation is
used to scale the theoretical sensitivity image to estimate the local
image RMS noise.  Table~2 lists the theoretical RMS
noise of the four field types defined in $\S2.1$: the raw mosaic image
RMS noise agrees with the predictions to within $\pm$ 20\%.

\begin{figure}
\begin{center}
\noindent\includegraphics[width=.45\textwidth]{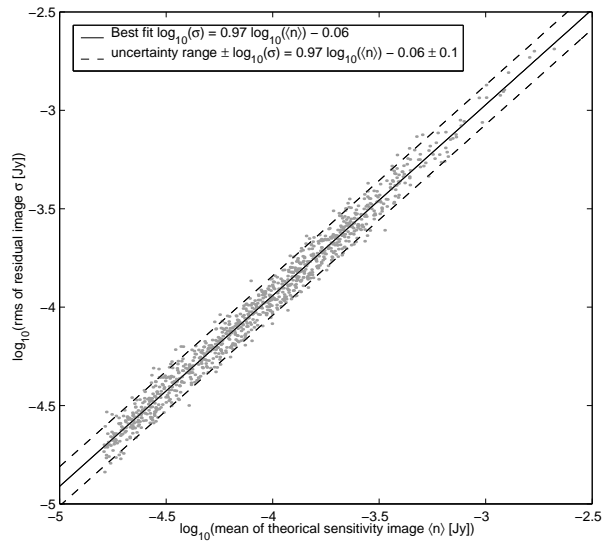}
\caption{Plot of residual image rms against theoretical sensitivity
image mean, with a best-fitting line (solid) and approximated range of
uncertainty (dashed-lines).} 
\label{fig:localnoise}
\end{center}
\end{figure}


\section{Radio detection of 2dFGRS galaxies}
In measuring the radio properties of 2dFGRS galaxies in our pilot 
survey, we restricted our radio image to a lowest RMS sensitivity 
of 1~mJy, the outermost contour in Fig.~3.
This is necessary as to limit the uncertainty in the primary beam 
corrections of our radio image. The total area to a 1~mJy detection
limit is $\sim$ 3 deg$^2$. Fig.~5 shows the
variation of survey area as a function of $1\sigma$ sensitivity.

\begin{figure}
\begin{center}
\noindent\includegraphics[width=.45\textwidth]{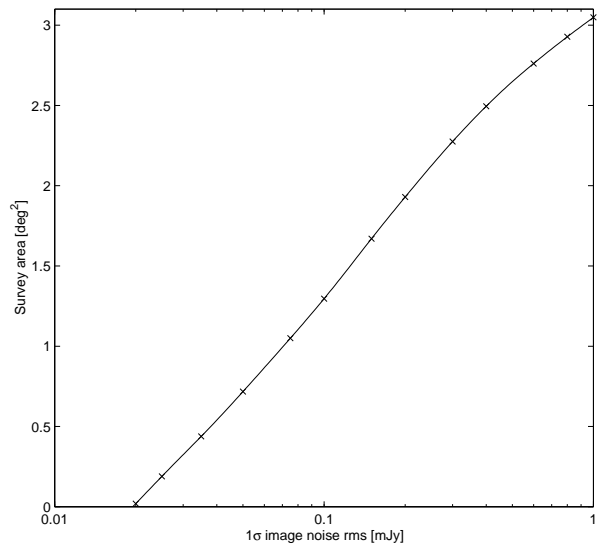}
\caption{Survey area as a function of $1\sigma$ RMS noise level. The
solid line is a cubic spline interpolation of the data.}
\label{fig:surveyarea}
\end{center}
\end{figure}

Based on the final 2dFGRS data release, a total of 392 2dFGRS objects 
lie within the 1~mJy-region of the ATCA mosaic. Of these, 27 have reliable 
(Q $\geq$ 3) low-redshift spectra ($z \leq 0.001$), they are stars and 
have been excluded from further analysis. The spectra of the remaining 365 
objects have been visually classified by author BC and cross-checked 
by author EMS. We adopted the visual classification scheme used by
\citet{Sadler02} a subset of which were also classified by 
\citet{Jackson00} using diagnostic emission-line ratios.  The spectra were
classified as AGN galaxies with (1) \textit{abs}: absorption line
spectra, (2) \textit{em}: emission-line spectra with dominant
forbidden lines (e.g. [OIII]), or (3) \textit{a+e}: combined
absorption and emission line spectra, or as star-forming galaxies with
(4) \textit{SF}: emission-line spectra with dominant recombination
lines (e.g. H$\alpha$), or as un-classifiable objects (5)
\textit{???}: objects with spectra which are too noisy or otherwise
un-classifiable. Note that due to aperture effects of fiber spectroscopy, 
a spectra classified as \textit{abs} or \textit{a+e} may depend on the 
redshift of the galaxy \citep[e.g. see][their Fig.4 and $\S$3.4]{Sadler02}. 
Fig.~6 exhibits sample 2dFGRS spectra in classes (1)--(4) 
from top to bottom.

\begin{figure}
\begin{center}
\noindent\includegraphics[width=.45\textwidth]{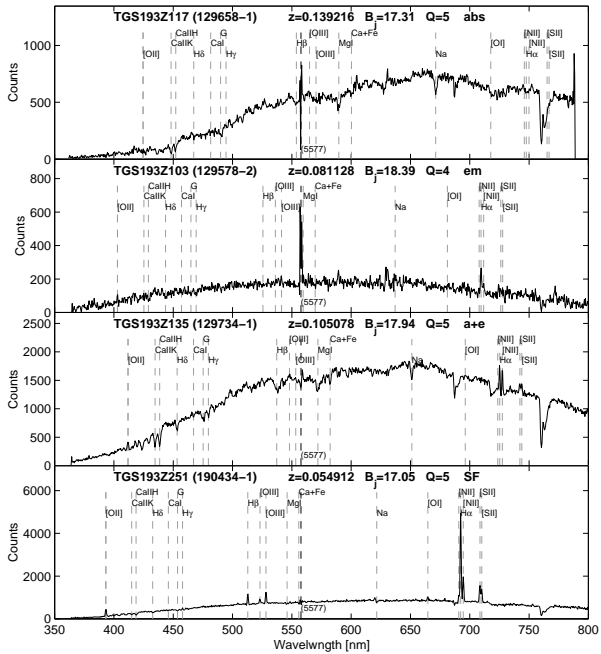}
\caption{Examples of 2dFGRS spectra, classified from the top as (1)
\textit{abs}, (2) \textit{em}, (3) \textit{a+e}, and (4) \textit{SF}.
See $\S$3 for details.}
\label{fig:egspectra}
\end{center}
\end{figure}

We aim to determine the radio flux density or an upper limit at each
2dFGRS source position. For the brighter radio sources, this can be
done by cross-matching a radio source catalogue to the 2dFGRS
catalogue. However, since the optical source positions are known we
can explore potential low signal-to-noise (S/N) detections by
examining the statistical outcomes of source fitting at the known
optical positions. Our strategy for cross-identification when the
optical positions are known employs the {\sc AIPS}
procedure {\sc vsad} to catalogue radio sources with peak flux density
$\geq$ 5 times the local image noise rms, and the {\sc MIRIAD}
procedure {\sc imfit} to examine the statistical properties of pixels
near the optical sources. Estimated uncertainty in the flux density 
measurements by the two source fitting programs have been derived using 
Monte-Carlo simulations, the results are presented in Appendix I.

We deploy {\sc vsad} for multiple Gaussian fitting within regions of
$\sim 3$ beam areas, centred in the previously determined CLEAN boxes
and adopting a peak flux density cutoff of 3 times the local RMS noise
determined as explained in $\S2.2$.  The resulting 3$\sigma$ peak flux
density list is filtered to produce a radio source list with peak flux
densities exceeding the local 5$\sigma$ level and inspected visually
to reject obvious spurious fits. This radio source catalogue,
containing 326 single components and 2 paired sources, is then 
cross-matched with the 365 2dFGRS positions. To determine the optimal 
search radius for cross-matches,
we consider Fig.~7 which shows the number of
matches (``$\circ$'' with $\sqrt{N}$ Poisson errorbars) inside
consecutive concentric annuli centred at the optical position. 
Monte-Carlo techniques were used to determine the level of chance 
coincidence in the case of variable survey sensitivity. The Monte-Carlo 
results (``$\ast$''-symbols) agreed well with the expected chance 
coincidence (dotted-line in Fig.~7) calculated for 
a uniform radio source density.  Fig.~7 shows that 
beyond a radio-optical offset of about 4-5 arcsec, the rate of
radio-2dFGRS cross-matches overlaps with that expected from chance
coincidence. Matches with offsets up to 5 arcsec are accepted as
automatic identifications, but we visually inspect matches up to 15 arcsec 
offset to uncover special cases.

Of the 365 2dFGRS positions, we found 23 identifications with radio-optical 
offsets smaller than 5 arcsec, and 3 positions with 5--10 arcsec offsets 
which were all accepted after visual inspection. The 2 matches with offsets 
10--15 arcsec were rejected as identifications.

\begin{figure}
\begin{center}
\noindent\includegraphics[width=.45\textwidth]{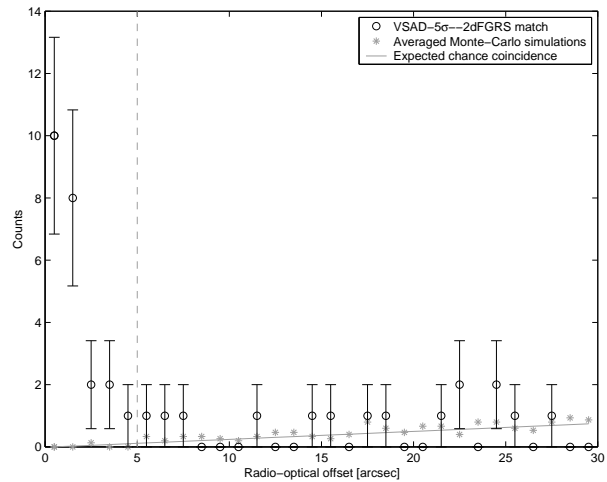}
\caption{The differential matches of radio-2dFGRS sources as a
function of radio-optical offset, shown with Poisson errorbars
($\sqrt{n}$). Also shown are averaged Monte-Carlo test results and the
solid line shown the expected chance coincidence level, assuming radio
sources are uniformly distributed.} 
\label{fig:searchradius}
\end{center}
\end{figure}

For each 2dFGRS position without a radio match, the {\sc miriad} task 
{\sc imfit} was used to attempt a least-squares fit of the radio point
spread function at the optical position.  Although the fitting
algorithm is understandably unstable, fits converged for 65 cases out 
of the remaining 339 optical positions, with a fitted peak flux density 
greater than the estimated local RMS noise ($1\sigma$). Of these, 9 
having a peak flux density $\geq 3\sigma$ are reported as flux density 
measurements. The 56 with fits below $3\sigma$ are characterised as 
upper limits. For the remaining 274 optical positions with unsuccessful 
{\sc imfit} results (divergent fits or fitted flux density $\leq 1\sigma$) 
we report a flux density of zero, and specify the estimated local noise as 
the radio upper limit.

We developed a \textit{glish}-script to generate image overlays using
methods in the {\sc viewer}-tool of {\sc AIPS++}.  Red 2$^{\rm nd}$ epoch
Digitized Sky Survey (DSSII-red) images\footnote{DSS images obtained
from the \textit{Canadian astronomy data centre}, at:
\textsf{http://cadcwww.dao.nrc.ca/dss/}} were overlaid with radio
contours at each of the 365 2dFGRS objects. These images led us to
accept all 3 matches with radio-optical offsets between 5--10 arcsec,
and also revealed two objects whose radio emission was extended and
required pixel summation to determine the integrated flux densities.
DSSII-red images of these objects are displayed in
Fig.~8, overlaid with radio contours.  The
dashed-outline in Fig.~8 shows the region used to
determine the radio flux density and the cross marks the corresponding
2dFGRS object position.

\begin{figure}
\begin{center}
\noindent\includegraphics[width=.45\textwidth]{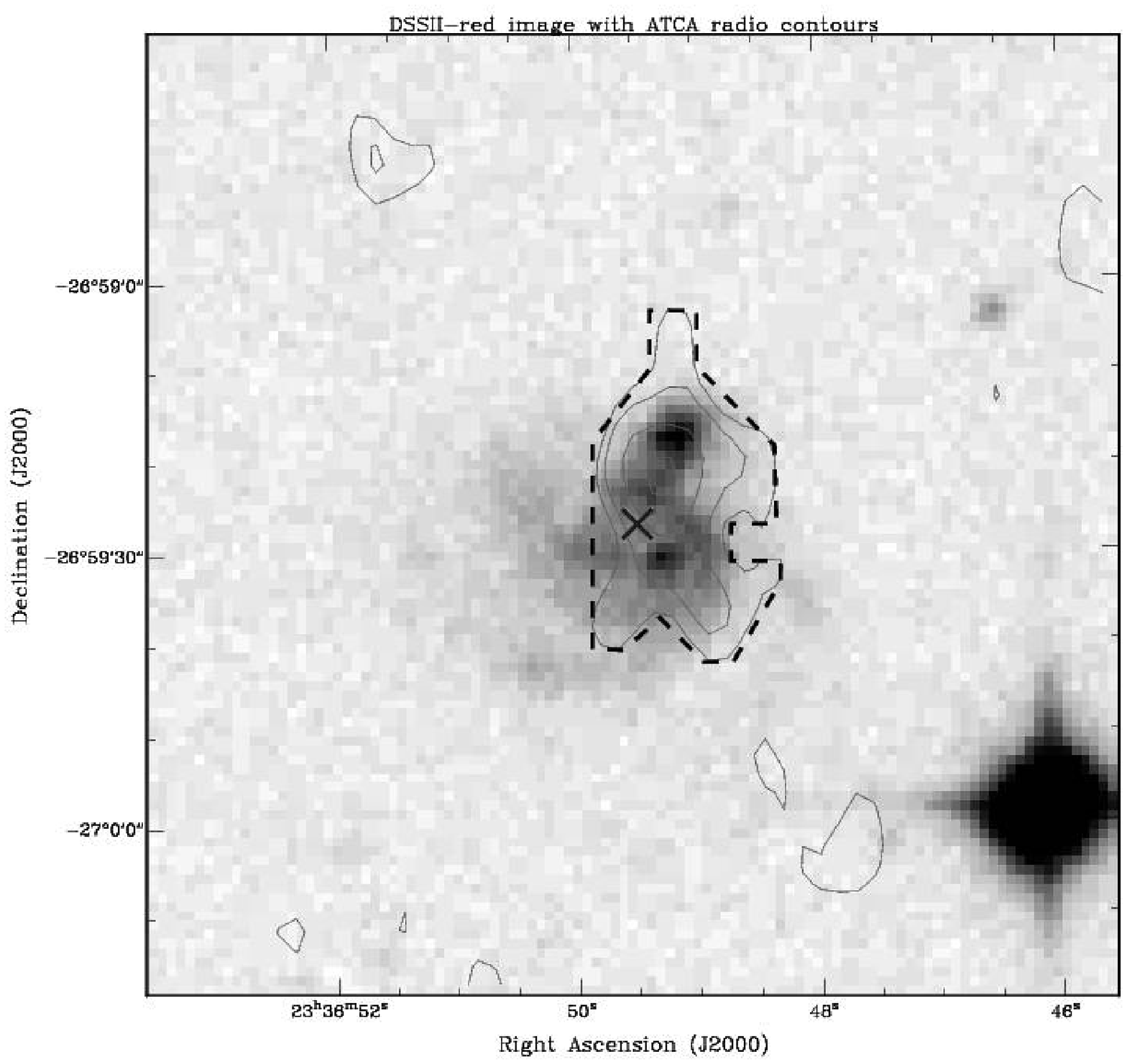}
\noindent\includegraphics[width=.45\textwidth]{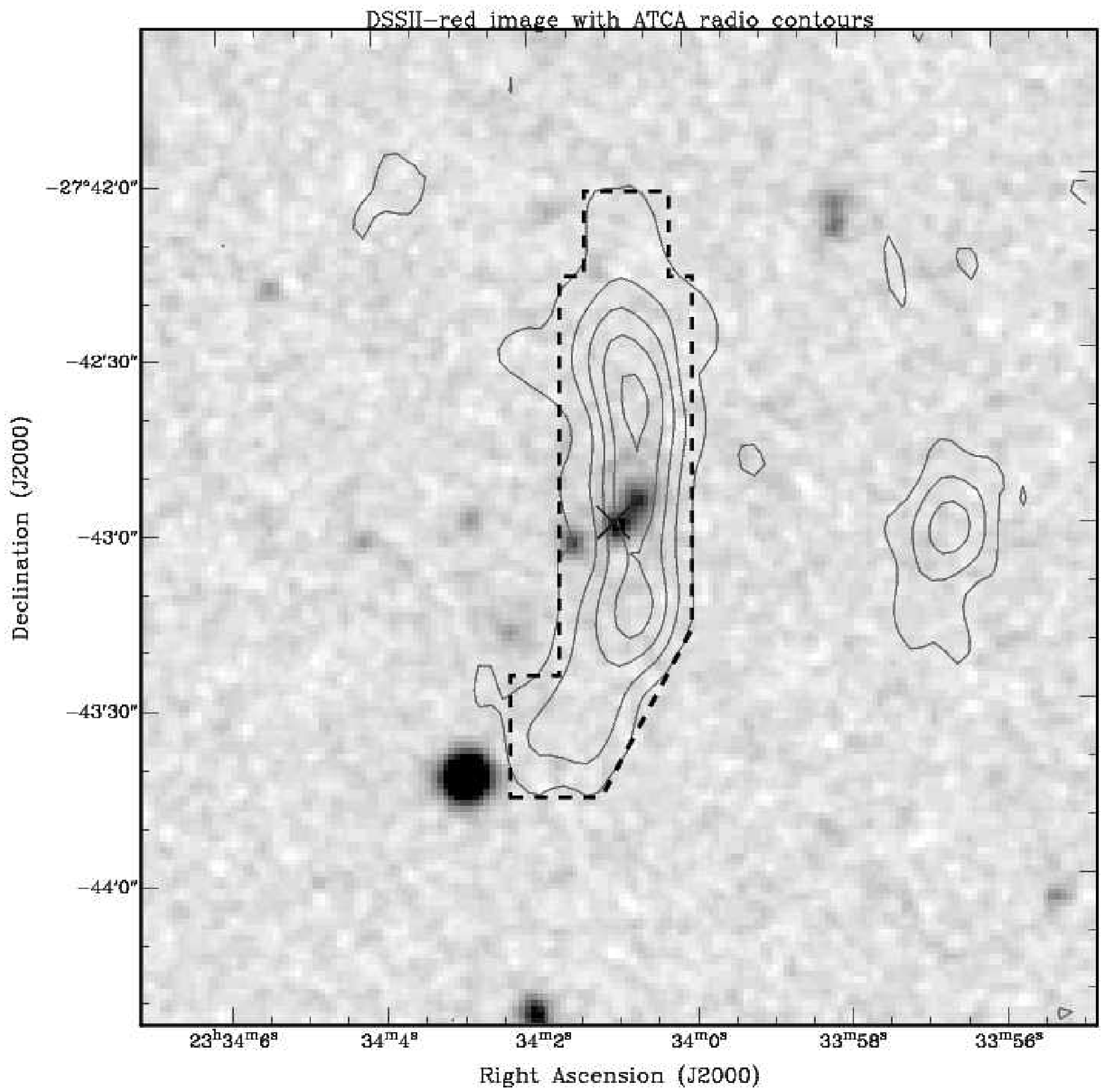}
\caption{ DSSII-red image overlaid with (top) radio contours at 2, 3,
5, and 7 $\times \sigma = 474~\mu$Jy) and (bottom) radio contours at
2, 6, 11, 17, and 24 $\times \sigma = 358~\mu$Jy).  The dashed outline
marks the region used to determine the radio flux density and the
cross marks the 2dFGRS position.}
\label{fig:pixelsum}
\end{center}
\end{figure}

\subsection{ATCA-2dFGRS data-table}
We publish here the radio flux density measurement or upper-limits of the 
365 2dFGRS objects within our ATCA radio survey region. Table 3 shows page~1 
of the full data table. This catalogue is sorted on right ascension, and 
the structure of the table is as follows:\\
\textit{Columns 1 \& 2}: 2dFGRS position (J2000);\\
\textit{Column 3}: 2dFGRS object name;\\
\textit{Column 4}: 2dFGRS redshift;\\
\textit{Column 5}: final ${\rm B_j}$ magnitude used in 2dFGRS;\\
\textit{Column 6}: 2dFGRS redshift measurement quality;\\
\textit{Column 7}: visual spectral classificaton;\\
\textit{Column 8}: fitting method used to determine the radio flux;\\
(1) {\sc vsad}-5$\sigma$ match;\\
(2) {\sc imfit} $\geq 3\sigma$ match;\\
(3) {\sc imfit} $1-3~\sigma$ match;\\
(4) local noise upper limit;\\
(5) pixel sum;\\
\textit{Columns 9 \& 10}: {\sc vsad}-5$\sigma$ fitted radio position;\\
\textit{Columns 11}: estimated {\sc vsad} fitted positional uncertainty in 
arcsec, see Appendix I;\\
\textit{Columns 12}: radio-optical angular separation;\\
\textit{Columns 13 \& 14}: peak \& integrated radio flux density determined 
using the corresponding fitting method;\\
\textit{Columns 15 \& 16}: percentage \& actual radio flux density 
uncertainty\footnote{When the local noise exceeds the fitting uncertainty, 
the larger value was used.}, see Appendix I;\\
\textit{Column 17}: estimated local noise, see $\S2.2$;\\
\textit{Column 18}: radio signal-to-noise ratio of the detection.\\

\subsection{Summary and comparison of results}

Of the 365 2dFGRS objects with $z > 0.001$, 316 have reliable 
spectral classification, with 176 and 140 classified as AGNs 
and star-forming galaxies, respectively. The remaining 49 objects 
were unclassified, having either a low quality flag (Q $<$ 3) or 
noisy spectra.  

Of the 35/365 (9.6\%) radio detections (measured radio flux density 
exceeding 3$\sigma$) of these galaxies, 12 were
classified as AGNs, 20 as star-forming galaxies, and 3 were
unclassified.  Table~4 presents a summary of the
radio detection rate of AGNs and star-forming galaxies in 2dFGRS for
the current survey and those identified with NVSS sources by
\citet{Sadler02}. It is difficult to compare the NVSS and ATCA values
in view of the varying sensitivity in the ATCA survey, but they do
reveal a rise in the detection rate with deeper radio sensitivity,
especially for the star-forming population.

Fig.~9 shows the redshift and radio
luminosity distributions of the radio-identified 2dFGRS sources
associated with AGN and star-forming populations.  To facilitate
comparison, each distribution were normalised by the total of the
distribution.  Note that the narrow range of radio luminosities found
in the ATCA sample is a selection effect owing to the combination of 
optical-radio survey cutoffs and the relatively small
surveyed area.  The sub-mJy ATCA survey detects an increasing
proportion of lower-luminosity AGNs lying over a redshift range
similar to that of sources detected above 2.7~mJy, as well as an
increasing proportion of higher-redshift star-forming galaxies having
a mean luminosity slightly lower than the NVSS sample. This implies
that a sub-mJy radio survey of the 2dFGRS will better reveal the
population in the faint end of the AGN and star-forming galaxy radio
luminosity function, and enable more reliable determination of the
evolution of the star-forming population to $z \approx 0.1$ (radio 
limited, 50~$\mu$Jy) and the AGN population to $z \approx 0.3$ (optical limited).

\setcounter{table}{3}
\begin{table}
\begin{center}
\caption{ATCA-2dFGRS summary.} \label{tab:matchsummary}
\vspace{1em}
\begin{tabular}{lrrr} \hline
				& 2dFGRS & ATCA & NVSS \\ 
				& & detection & detection \\
				\hline z-unreliable & 49 & 3 (6.1\%) &
				--- \\ AGN & 176 & 12 (6.8\%) & $\sim$
				1.8\% \\ SF & 140 & 20 (14\%) & $\sim$
				1.1\% \\ Total & 365 & 35 (9.6\%) &
				$\sim$ 1.5\% \\ & & & \\
				\multicolumn{2}{l}{AGN:z$_{\rm med}$
				(sd)} & 0.14 (0.046) & 0.14 (0.16) \\
				\multicolumn{2}{l}{SF:z$_{\rm med}$
				(sd)} & 0.086 (0.037) & 0.046 (0.040)
				\\ & & & \\
				\multicolumn{2}{l}{AGN:log$_{10}(L_{1.4~GHz})_{\rm
				med}$ (sd)} & 22.22 (0.69) & 23.58 (0.81) \\
				\multicolumn{2}{l}{SF:log$_{10}(L_{1.4~GHz})_{\rm
				med}$ (sd)} & 21.93 (0.47) & 22.38 (0.58) \\ \hline
\end{tabular}\\
\end{center}
\end{table}

\begin{figure}
\begin{center}
\noindent\includegraphics[width=.45\textwidth]{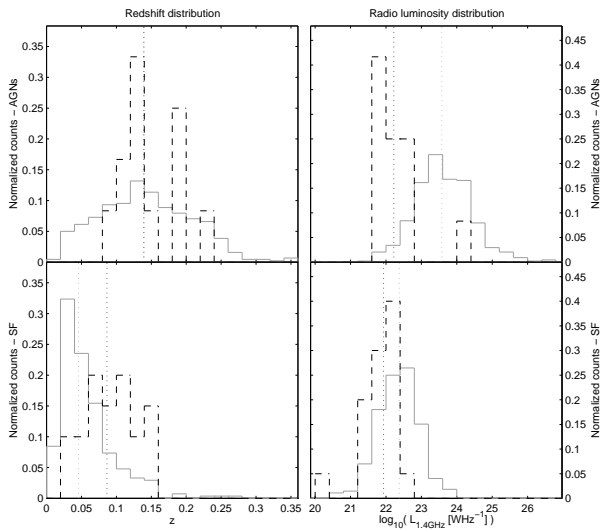}
\caption{Redshift (left-column) and radio luminosity (right-column)
distributions for AGNs (top-row) and star-forming galaxies
(bottom-row) of 2dFGRS sources identified with this current radio
survey (dashed-line-histogram) and NVSS sources (solid-line-histogram)
from \citet{Sadler02}. The two vertical dotted-lines in each plot
shows the median of the distributions, the actual values are listed in
Table~4.} 
\label{fig:ATCA_NVSS_distributions}
\end{center}
\end{figure}

\section{Radio luminosity function}
In this section we describe the construction of the total radio
luminosity function (RLF) and the corresponding contributions by the
AGNs and star-forming (SF) populations for our ATCA-2dFGRS sample. We
have used the traditional 1/$V_{max}$ method of \citet{Schmidt68} to
derive the RLF, as described in $\S2$ of \citet{Condon89}. According
to the combined optical and radio survey constraints we calculate, for
each source, the maximum detectable redshift range
$[z_{min},z_{max}]$.  Adopting the Einstein-de-Sitter model, the
maximum detectable co-moving volume ($V_{max}$) is given by:

\begin{equation} 
V_{max}=16\pi(\frac{c}{H_o})^3\int_{z_{min}}^{z_{max}}\omega(z)\frac{[1-(z+1)^{-1/2}]^2}{(z+1)^{3/2}}dz,
\label{equ:co_V} 
\end{equation} 

where $\omega(z)$ is the fraction of
the surveyed solid angle at redshift $z$. The adopted optical survey
limit is 14.0 $\leq b_J \leq$ 19.4. (\citet{Colless01}), constant
throughout the survey. The radio detection limit varies across the
radio mosaiced region (see Fig.~3), and is set at
3$\sigma$.  A cubic spline interpolation of the relation shown in
Fig.~5 provides an estimate of $\omega(z)$ for any
particular source, and $V_{max}$ is obtained via numerical integration
of equation~\ref{equ:co_V} following the method described in $\S2$(d)
of \citet{Avni80}. Similarly, we can calculate the detectable co-moving 
volume of each source ($V$). The ratio $V/V_{max}$ of a randomly 
distributed sample is expected to be uniform in the interval [0,1], 
and the mean ($\langle V/V_{max} \rangle$) should be $\sim$ 0.5. 

Our sample comprises 35 ATCA-2dFGRS sources which satisfy the survey
limits stated above.  Three without reliable redshifts are excluded
from the RLF calculations, but included through the incompleteness
correction.  To estimate the level of incompleteness of our
ATCA-2dFGRS RLF, we examined galaxies in the 2dFGRS input catalogue
that fall within the radio surveyed region but were not observed with
2dF. Only one of these ``missing'' 2dFGRS sources was identified with
a $\geq 3\sigma$ radio source, implying that this optically selected 
radio-sample is approximately 90\% (32/36) complete.

The total RLF and contributions by the AGNs and SF galaxies are listed
in Table~5, also listed are the $\langle V/V_{max} \rangle$ 
values. While the SF galaxies in our sample seems to be uniformly 
distributed, the $V/V_{max}$ values of the AGNs show a small but not 
significant increase and they are non-uniformly distributed 
(median($V/V_{max}$) = 0.617). This might indicate some level of 
clustering or evolution in the AGN fraction within our sampled volume.

Presented in Fig.~10 is the
total RLF of our ATCA-2dFGRS sample ($\ast$-solid) along with values
derived from the NVSS-2dFGRS sample of \citet{Sadler02}. The RLF
separated into the populations of AGNs and SF galaxies of the two
samples is shown in Fig.~11.  The values and the plots
have not been corrected for the 10\% incompleteness since the effect
($\log_{10}(1.1)$) is small. Our RLF agrees well with that of 
\citet{Sadler02}, within measurement uncertainty and the luminosity 
range sampled, due to the smiliar redshift range sampled. Although 
our median redshift for the star-forming galaxies 
are higher, with such a small sample we are not yet able to test 
for any significant differences with respect to other samples.

\begin{table}
\begin{center}
\caption{ATCA-2dFGRS radio luminosity function.} \label{tab:RLF}
\vspace{1em}
\begin{tabular}{crcrcrc} \hline
& \multicolumn{2}{c}{All galaxies}	& \multicolumn{2}{c}{SF galaxies}	&	\multicolumn{2}{c}{AGNs} \\ 
$log_{10}(L^\diamondsuit)$	& N	& log$_{10}(\rho^\heartsuit_{\rm m})$ & N	& log$_{10}(\rho^\heartsuit_{\rm m})$ & N	& log$_{10}(\rho^\heartsuit_{\rm m})$\\ \hline 
20.5	&  1	& -2.62$^{+0.30}_{-1.00}$   &  1	& -2.62$^{+0.30}_{-1.00}$   &  -	& -                       \\
21.3	&  5	& -3.40$^{+0.22}_{-0.48}$	&  5	& -3.40$^{+0.22}_{-0.48}$	&  -	& -                       \\
22.1	& 23	& -3.50$^{+0.09}_{-0.11}$	& 14	& -3.67$^{+0.11}_{-0.15}$	&  9	& -3.99$^{+0.13}_{-0.19}$ \\
22.9	&  2	& -4.62$^{+0.24}_{-0.60}$	&  -	& - 						&  2	& -4.62$^{+0.24}_{-0.60}$ \\
23.7	&  0	& $\leq$-5.49				&  -	& - 						&  0	& $\leq$-5.49             \\
24.5	&  1	& -5.76$^{+0.30}_{-1.00}$	&  -	& - 						&  1	& -5.76$^{+0.30}_{-1.00}$ \\ \hline
Total   & 32    &                           & 20    &                           & 12    &                         \\
$\langle V/V_{max} \rangle$ & \multicolumn{2}{c}{0.526 $\pm$ 0.051} & \multicolumn{2}{c}{0.492 $\pm$ 0.065} & \multicolumn{2}{c}{0.584 $\pm$ 0.083} \\ \hline
\end{tabular}\\
\end{center}
$^\diamondsuit$ $L_{\rm 1.4GHz}$ in [WHz$^{-1}$]\\
$^\heartsuit$ $\rho_{\rm m}$ in [${\rm mag^{-1}Mpc^{-3}}$]\\
\end{table}

\begin{figure}
\begin{center}
\noindent\includegraphics[width=.45\textwidth]{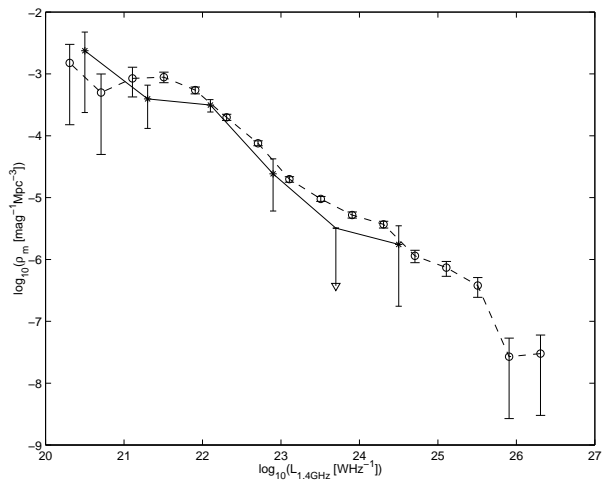}
\caption{Total radio luminosity functions for the 2dFGRS-NVSS sample of 
\citet{Sadler02} ($\circ$ dashed-line) and our 2dFGRS-ATCA sample 
($\ast$ solid-line).} 
\label{fig:RLFtotal}
\end{center}
\end{figure}

\begin{figure}
\begin{center}
\noindent\includegraphics[width=.45\textwidth]{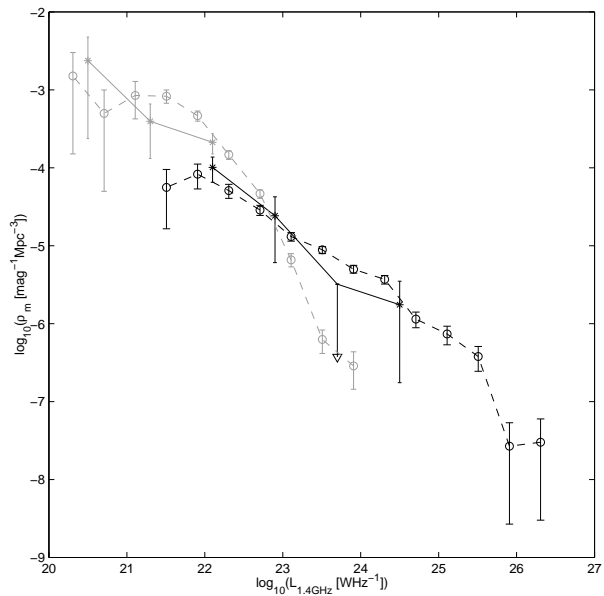}
\caption{Radio luminosity functions separated into the star-forming
(lighter-colour lines) and the AGN (darker-colour lines) populations
for the NVSS ($\circ$ dashed-line) and ATCA ($\ast$ solid-line)
samples.} 
\label{fig:RLFsep}
\end{center}
\end{figure}

\section{Radio-2dFGRS detection rate}

To explore the radio-2dFGRS detection rate at various radio detection
limits, the radio survey was divided into four zones of sensitivity,
starting from 25$\mu$Jy and doubling in four steps upwards.  The four
zones thus corresponded to $3\sigma$ radio detection limits of 75,
150, 300 and 600~$\mu$Jy.  The number of 2dFGRS sources and radio
detections above the $3\sigma$ limit are counted in each zone,
separated into three optical spectral classes: (i) star-forming (SF)
galaxies, (ii) AGNs and (iii) no redshift (unknown).  These counts are
listed in Table~6, excluding the detections above the 
2.8~mJy limit used in \citet{Sadler02}. 

Our survey region has been specially selected to avoid bright radio 
sources, therefore the radio-2dFGRS detection rates calculated from our 
sample could be biased against identifications with bright sources.
To overcome this possible bias, we have adopted the expected detection 
rate for sources above 2.8~mJy using the NVSS-2dFGRS results of 
\citep{Sadler02}. 
Fig.~12 illustrates the proportion of 2dFGRS sources identified with 
radio sources as a function of radio survey sensitivity. This is 
calculated from the sum of the detection rates at the various radio 
detection limits listed in Table~6 and the 
corresponding detection rate from the NVSS-2dFGRS sample.

The plot shows an approximate
10-fold increase in the detection rate for a 10-fold improvement in
radio detection sensitivity. This increase in detection rate would
require a 100-fold increase in observing time. 
These results are broadly consistent with the estimates made in $\S2$, 
using our observing strategy, we can obtain a sample evenly distributed 
in numbers of identifications, independent of the additional constraint 
imposed by 2dFGRS optical selection, across all radio flux densities.

\begin{table*}
\begin{center}
\caption{Radio-2dFGRS identification counts at the corresponding 
$3\sigma$ radio limit up to the 2.8~mJy limit used in \citet{Sadler02}.} 
\label{tab:subsamples}
\vspace{1em}
\begin{tabular}{ccrrrrrrrrrr} \hline
Radio $3\sigma$ limit	& Area 		& \multicolumn{4}{c}{Number of 2dFGRS galaxies} & \multicolumn{4}{c}{Number of radio identifications} \\ 
(mJy)					& (deg$^2$)	& Total   & SF   & AGNs   & unknown   			& Total   & SF   & AGNs   & unknown                   \\ \hline 
0.075 -- 2.8			& 0.19		&  27     &  8   &  17    &  2                  &   7     &  2   & 5      & --                        \\
0.150 -- 2.8			& 0.72		& 100     & 43   &  45    & 12                  &  18     & 10   & 6      &  2                        \\
0.300 -- 2.8			& 1.30		& 163     & 73   &  74    & 16                  &  14     & 10   & 4      & --                        \\
0.600 -- 2.8			& 1.93		& 227     & 97   & 101    & 29                  &   6     &  4   & 2      & --                        \\ \hline
\end{tabular}\\
\end{center}
\end{table*}

\begin{figure}
\begin{center}
\noindent\includegraphics[width=.45\textwidth]{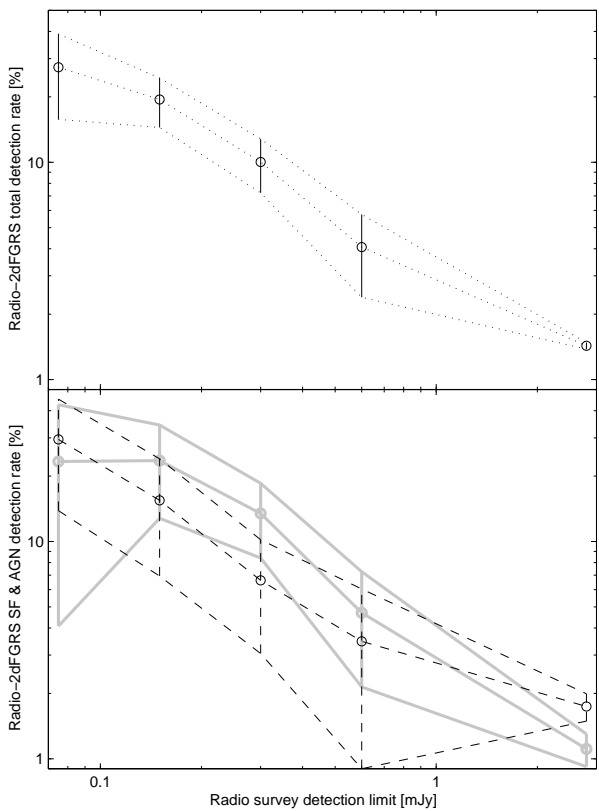}
\caption{Radio-2dFGRS detection rate at various radio detection
limits. The lines interpolates the detection rate with Poisson 
errors for: {\it top-panel} -- the total-detection rate (dotted-line);
{\it bottom-panel} -- detection rates of star-forming galaxies
(solid-line) and the AGNs (dashed-line). Note: (1) the detection 
rate at 2.8~mJy is from the NVSS-2dFGRS sample of \citet{Sadler02}, 
(2) the detection rates above 2.8~mJy are the {\it sum} of 
the corresponding NVSS-2dFGRS detection rate and the values shown 
in Table~6, see text for details.}
\label{fig:detectionrate}
\end{center}
\end{figure}


\section{Large scale radio survey}
Owing to the enormous time investment in acquiring large optical-spectroscopic 
redshift surveys such as that by 2dF makes it even more valuable to fully 
exploit its potentials. A large scale radio follow-up survey of the 2dF 
sources will fill a significant gap in the available data on faint radio 
sources. Existing radio-optical spectroscopic studies have been based 
on either (i) bright, large-scale radio surveys (FIRST\footnote{Very Large 
Array--Faint Images of the Radio Sky at Twenty-centimeters \citep{Becker95}.}, 
NVSS and SUMSS\footnote{Sydney University Molonglo Sky Survey 
\citep{Bock99}.}) that are cross-correlated with catalogued optical data 
(implying that the identified optical galaxies are also bright) or (ii) deep 
radio surveys made over small areas (such as: PDS\footnote{Phoenix Deep Survey 
\citep{Hopkins98}.}, HDF\footnote{Hubble Deep Field \citep{Richards98}.} 
and HDFS\footnote{Hubble Deep Field South \citep{Norris01}}) with pointed 
optical spectroscopy of selected optical IDs. Studies of type (i) have 
given radio-optical data on $\sim 4000$ and $\sim 1000$ nearby galaxies 
from the UGC-NVSS \citep{Condon02} and 2dF-NVSS \citep{Sadler02} \& 
2dF-FIRST \citep{Magliocchetti02}, with $z <$ 0.1 and $z <$ 0.3, 
respectively. Studies of type (ii) (which is very expensive of telescope 
time) has to date provided radio-optical data on fewer than 300 galaxies, 
at redshifts up to $z \sim$1. 

Fig.~13 displays the radio-detected 2dFGRS objects in
the current survey and those in 2dF-NVSS. Also shown are dashed lines 
tracing out the UGC-NVSS sample, and a solid line delineating the main 
concentration of UGC-NVSS data points (predominately star-forming 
galaxies). It is clear that mJy radio surveys are only capable of 
properly sampling the full range of radio luminosities for the SF and 
AGN populations up to $z \sim$ 0.01 ($L_{\rm 1.4GHz} \gta 10^{21.5}$~WHz$^{-1}$). 
Whereas, even though the ultra-deep 
radio surveys can provide radio luminosity sampling to very high redshifts 
($z \geq$ 1), the problems lies in that pencil beam surveys (i.e. small 
survey area of typically a few square arcmin up to a few square degrees) 
are not ideal to sample galaxy evolution. There are inherent problems in 
small area surveys associated with clustering, cosmic variance, and the 
complex selection effects will be both difficult to model and correct for.
A large area sub-mJy radio surveys of 2dFGRS sources with a varying 
sensitivity will link together the existing surveys and provide the 
required sample of the `normal' population of sources in the intermediate 
redshift ranges up to $z \sim$ 0.03-0.04. An investigation of possible 
systematic source flux differences between our ATCA survey and those in 
the NVSS, we present, in Appendix II, flux comparison of sources found 
common in the two surveys within our mosaiced region.

\citet{Sadler02} estimate that approximately 4,000 radio-optical
identifications will be found when the full 2dFGRS is correlated with
the NVSS catalogue (Jackson et al. 2003, in preparation).  Using the 
detection rates measured with our pilot study, we can now estimate the 
survey sensitivity and the total survey area (i.e. observing time) 
required to provide a set of radio IDs that is complete to a given 
radio luminosity and redshift. Computed in (Table~7) 
is the required observing time and the number of expected optical-radio 
identifications for a survey with three levels of sensitivity and survey 
area. The three adopted 3$\sigma$ sensitivity levels form a bridge 
between existing surveys, the poorest sensitivity level at 0.6~mJy 
being approximately half of the FIRST limit (the FIRST limit is 
approximately half of NVSS limit) while the best sensitivity level 
of 0.15~mJy is about half as good as that of PDS.

Employing a nested mosaicing technique the survey plan outlined in 
Table~7 will require a total of about 60--70 $\times$ 
12 hour observing sessions. The survey will image about 100~deg$^2$ 
of sky, making over 1000 radio/optical identifications. This number 
is required to address the
important problem of constraining the cosmic evolution of the radio
spectral power density of star-forming galaxies. The UGC-NVSS sample
has provided a very precise measurement of the local ($z \sim$ 0.01)
star formation density, presenting a local reference and has 
hinted at intriguing cosmic evolution of this quantity \citet{Condon02}. 
With favored evolutionary model in which the luminosity density rises 
as $(1+z)^{3\pm1}$, the predicted increase of star formation rate density
is $\sim$ (30 $\pm$ 10)~\% at $z \sim$ 0.1.  With a sample size of 1000,
divided into ten luminosity bins for deriving the redshift-dependence 
of the star formation density, producing a measurement uncertainty of 
$\sim$ 10\%. This would be sufficient in distinguishing between various 
models for the cosmic evolution of the star formation density, providing 
a measure potentially superior to the traditional estimates via optical 
emission-line measurements or ultra-violet photometry which are unavoidably 
biased by dust extinction.

\begin{table*} 
\centering
\caption{Sample observing plan.}
\label{tab:OBSplan} \vspace{1em} 
\begin{tabular}{lrrrrrrrr} 
\hline 
   (1)    &      (2)       &       (3)    &     (4)     &      (5)     &       (6)      &        (7)    &     (8)      &    (9)   \\
  Area    &   \# pointing  &    3$\sigma$ &   ID rate   &   time/field &    \# of 12hr  &  \#ID/deg$^2$ &  \#ID below  &   Total  \\ 
(deg$^2$) &                &      (mJy)   &     (\%)    &     (min)    &    observation &               &   NVSS/deg   &    \#ID  \\ \hline 
98.0      &      1221      &      0.60    &       5     &      7.5     &        13      &         5.12  &     3.6      &     351  \\ 
49.0      &       610      &      0.30    &      10     &     30.0     &        25      &        10.23  &     8.7      &     426  \\ 
24.5      &       305      &      0.15    &      20     &    120.0     &        51      &        20.47  &    18.9      &     464  \\ \hline 
\end{tabular}\\ 
\begin{tabular}{l}
(1) Area surveyed (in square degrees) to specified sensitivity.\\
(2) Number of pointings needed to cover required area.\\
(3) Radio 3$\sigma$ (3 $\times$ image noise rms) sensitivity limit.\\
(4) Percentage of expected optical-radio identification rate, obtained from pilot study.\\
(5) Amount of radio telescope time required to reach sensitivity limit per pointing.\\
(6) Number of 12 hour observing sessions required to cover area.\\
(7) Number of optical-radio identification expected per square degree.\\
(8) Number of optical-radio identification expected per square degree inaccessible to NVSS.\\
(9) Total number of optical-radio identification down to the sensitivity limit inaccessible to NVSS.\\
\end{tabular}
\end{table*}

\begin{figure}
\begin{center}
\noindent\includegraphics[width=.45\textwidth]{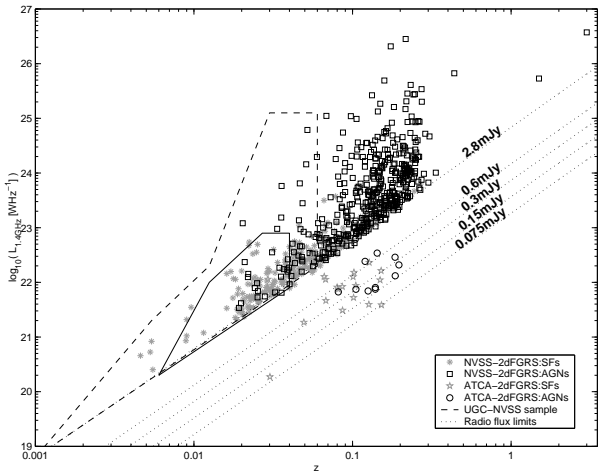}
\caption{Plot of radio luminosity against redshift, for our 
ATCA-2dFGRS (3$\sigma$) detections along with the NVSS-2dFGRS 
($\sim 5\sigma$) sample. The dashed-line traces out the UGC-NVSS 
($\sim 5\sigma$) sample, the solid-line shows the main concentration of 
data points (predominately star-forming galaxies). The dotted-lines 
shows various flux density limits.}
\label{fig:Lzplot}
\end{center}
\end{figure}

\section*{Acknowledgments} 
We thank the 2dFGRS team for data access prior to the public release.

We thank R.J. Sault for helpful advice
regarding data reduction and joint deconvolution techniques.  We also
wish to thank M. Marquarding for helpful demonstration of the {\sc
AIPS++ viewer}-tool functionalities and answering many queries with
respect to \textit{glish}-scripting.

The Australia Telescope is funded by the Commonwealth of Australia for
operation as a National Facility managed by CSIRO.

\section*{Appendix I: Source fitting and uncertainties}
In measuring the radio flux density associated with 2dFGRS sources, we have used 
two automatic radio source fitting packages: (1) the {\sc aips} task {\sc vsad}, 
and (2) the {\sc miriad} task {\sc imfit}. We first matched the optical source 
list to a radio catalog generated with {\sc vsad}, of bright ($\geq$ 5$\sigma$) 
radio sources. For positions without a bright radio counterpart, we used 
{\sc imfit} to determine radio upper-limits by fitting a beam size elliptical 
Gaussians at the known optical positions. Following these techniques, we 
investigated the source fitting uncertainty with Monte-Carlo simulations. 

Beam-size Gaussians were injected at randomly chosen positions into the residual 
image of the deep-CLEANed mosaic. To obtain an estimate of the source fitting 
accuracy as a function of fitted source strength\footnote{The fitted source strength 
($\sigma_s$) is defined as the ratio of the fitted peak flux density ($S_{peak}$) to 
the estimated local noise ($\sigma_n$, see $\S 2.2$).} ($\sigma_s$), a range of 
injected source strengths were used, corresponding to a signal-to-noise ratio up to 
$150$ $\sigma$. We first used {\sc vsad} to generate a 
5$\sigma$ radio source list, following the same procedures as we used in our study 
(see $\S$3). The input position list is then matched against the 5$\sigma$ radio 
source list to within 10 arcsec, while fitting with {\sc imfit} at the known source 
position for those without a match. 

Shown in Fig.~14 is the Monte-Carlo results for the {\sc vsad} 
fitted angular positional deviation from the input at various fitted source 
strengths. The solid-curve is a quadratic least-square fit to the logarithmic 
one-standard deviation binned values of the results. This curve bounds about 90\% 
of the data points. For instance, for about 90\% of the 5$\sigma$ {\sc vsad} 
sources in our radio catalog, we expect to have a positional uncertainty of 
$\lta$ 3 arcsec. 

The results of the Monte-Carlo simulations for the two source fitting packages 
are shown in Fig.~15. Plotted in linear-log scales is the 
fitted peak flux fractional difference $dS/S = (S_{input}-S_{fitted})/S_{fitted}$, 
as a function of source strength. The two solid curves are quadratic least-square
fits to the one-standard-deviation level of $\log_{10}(|dS/S|)$. 
The upper and lower bounds were fitted separately to better represent the complex 
behavior of the source fitting algorithm in the presence of noise. The uncertainty 
has been set to a absolute minimum of 5\% for strong sources. Overall, the 
estimated uncertainty bounds $\gta$ 90\% of the data. The deficit 
of data points with low fitted flux densities ($dS/S < 0$) at faint source strengths 
is the well-known property of radio source fitting in the presence of noise, as 
described in \citet{Condon97}.

For the {\sc vsad} results in our present work, we 
have made the assumption that the fractional uncertainty in fitted peak flux 
density is representative of the fitted integrated flux density.

\begin{figure}
\begin{center}
\noindent\includegraphics[width=.45\textwidth]{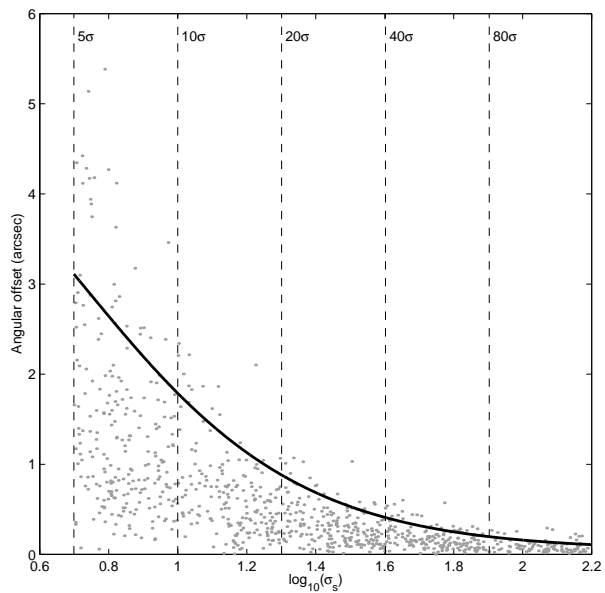}
\caption{The angular separation between the input and fitted source position as 
a function of fitted source strength, in linear-log scales. The solid curves 
indicates the fitted estimate of the positional fitting uncertainty as a 
function of fitted source strength. The vertical dashed-lines denotes the various 
fitted signal-to-noise ($\sigma$) level.}
\label{fig:mcseptest}
\end{center}
\end{figure}

\begin{figure}
\begin{center}
\noindent\includegraphics[width=.45\textwidth]{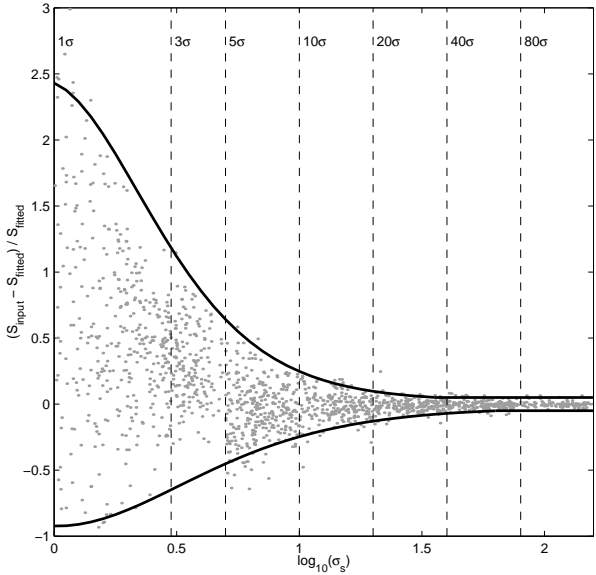}
\caption{The fitted peak flux deviation as a function of fitted source strength 
for the Monte-Carlo simulations, in linear-log scales. The solid curves 
indicates the fitted estimate of the fractional source fitting uncertainty as a 
function of fitted source strength. The vertical dashed-lines denotes the various 
fitted signal-to-noise ($\sigma$) level.}
\label{fig:mcfluxtest}
\end{center}
\end{figure}


\section*{Appendix II: ATCA and NVSS flux scale}
Radio sources have an extremely wide range of luminosities and usually requires 
a combination of surveys at various sensitivity and aerial coverage to properly 
sample the radio population. In this section, we consider the possibility of 
combining radio data from a survey with the ATCA and the NVSS from the VLA. We 
compared the source flux density measured in our survey and those in NVSS. With the 
on-line NVSS catalog browser\footnote{http://www.cv.nrao.edu/nvss/NVSSlist.shtml} 
we extracted the 157 NVSS sources, brighter than 2.5~mJy, lying within our current 
ATCA-2dFGRS mosaic region. NVSS has a resolution of 45 arcsec and this is typically 
the minimum angular separation between sources in the NVSS catalog. With higher 
angular resolution, some NVSS sources are resolved in our ATCA survey. To locate 
these `decomposed' sources, we matched the 157 NVSS positions to our 5$\sigma$ 
{\sc vsad} list using a search radius of 45 arcsec, centred at the NVSS source 
positions. For matches with multiple components we represent the ATCA flux by simply 
adding the fitted integrated flux of the components. We used the NVSS flux uncertainty 
(provided with the catalog) and our estimated flux uncertainty (derived with Monte-Carlo 
simulations, see Appendix I) as a measure of the degree of deviation (the ``sigma-level'') 
of the data from the prefect one-to-one correlation. For instance, a source with NVSS flux 
$S_{\rm NVSS} \pm \Delta_{\rm NVSS}$ and corresponding ATCA flux $S_{\rm ATCA} \pm 
\Delta_{\rm ATCA}$, we defined the deviation ``sigma-level'' ($\Delta_\sigma$), such that, 
for $S_{\rm NVSS} > S_{\rm ATCA}$, 
\[
S_{\rm NVSS} - \Delta_\sigma\Delta_{\rm NVSS} = S_{\rm ATCA} + \Delta_\sigma\Delta_{\rm ATCA},
\]
or in general, 
\[
\Delta_\sigma = \left| \frac{S_{\rm ATCA} - S_{\rm NVSS}}{\Delta_{\rm NVSS}+\Delta_{\rm ATCA}} \right|.
\]
We visually examined all NVSS positions with $\Delta_\sigma \geq$ 1.5 and those without 
an ATCA match. We found for a few extended sources, the Gaussian fitting method was 
inappropriate and we adopted the sum of the pixel values as the flux measurement. There 
are also a few cases where the source structure required some special care. The results 
of the cross identification process is listed in Table~8 and plotted 
in Fig.~16. From this small study, we did no observe any signs of 
significance systematic differences between the two radio surveys, in particular, there 
do not seem to be any systematic lost of total radio flux density in the ATCA survey for 
intrinsically extended sources (i.e. sources resolved by ATCA). 

\begin{table} 
\begin{center} 
\caption{ATCA--NVSS cross identification results.}
\label{tab:ATCANVSS} \vspace{1em} 
\begin{tabular}{lrrrrrrr} 
\hline 
Type  & $\Delta_\sigma$:\hspace{1em}$\leq$1 & 1--2  & 2--3 & $>$3 & Total & \\ \hline 
(1)   & 72  	  & 29  	  & 7		  & 0		 & 108 & (69\%)    \\ 
(2)   & 16  	  & 3		  & 0		  & 0		 & 19  & (12\%)    \\ 
(3)   & 1		  & 0		  & 0		  & 0		 & 1   & ($<$1\%)  \\ 
(4)   & 7		  & 2		  & 0		  & 1		 & 10  & (6\%)     \\ 
(5)   & 9		  & 0		  & 0		  & 0		 & 9   & (6\%)     \\ 
(6)   & 0		  & 0		  & 6		  & 4		 & 10  & (6\%)     \\ \hline
Total & 105 	  & 34  	  & 13  	  & 5		 & 157 & 		   \\ 
      & (67\%)    & (22\%)    & (8\%)	  & (3\%)	 &     &    	   \\ \hline
(7)   & 0		  & 6		  & 1		  & 2		 & 9   &           \\ \hline
\end{tabular}\\ 
\end{center} 
(1) Single ATCA component match.\\
(2) Double ATCA components match.\\
(3) Triple ATCA components match.\\
(4) Extended structure requiring pixel summation.\\
(5) Special cases.\\
(6) Undetected in ATCA.\\
(7) ATCA sources $\geq$ 2.5~mJy without NVSS match ($\sim$ 6\%).\\
\end{table}

\begin{figure}
\begin{center}
\noindent\includegraphics[width=.45\textwidth]{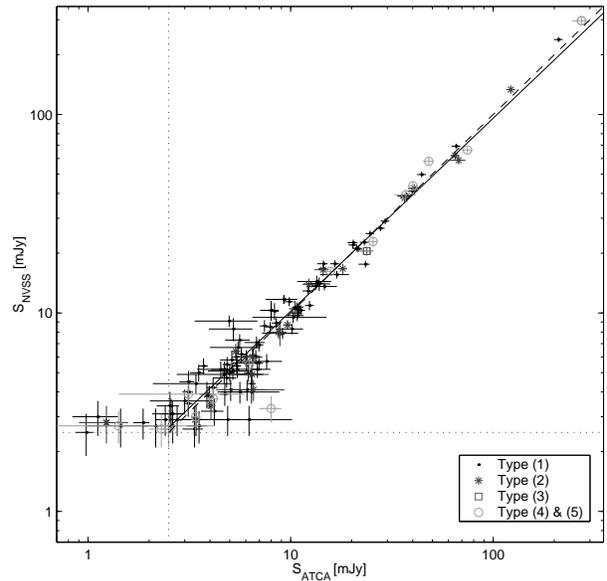}
\caption{The NVSS and ATCA flux measurements of sources matched within our survey region.
The different types of symbols denoted in the figure correspond to the matched-types 
listed in Table~8. The solid line is the least square fit to the data, 
the gashed-line is the one-to-one correlation between NVSS and ATCA fluxes. The dotted 
lines notes the 2.5~mJy NVSS limit adopted.}
\label{fig:ATCANVSS}
\end{center}
\end{figure}

\clearpage
\begin{onecolumn}
\begin{sidewaystable}
{\large {\bf Table 3.} Sample table of 2dFGRS-radio identifications and upper-limits.}\\
\tiny
\begin{tabular}{cccccccccccccccccc}
\hline
\multicolumn{2}{c}{2dF position (J2000)}                     & \multicolumn{1}{c}{2dF name}  & \multicolumn{1}{c}{z}  & \multicolumn{1}{c}{\rm $B_j$}  & \multicolumn{1}{c}{Q}  & \multicolumn{1}{c}{class}         & \multicolumn{1}{c}{FM$^\spadesuit$}	& \multicolumn{3}{c}{{\sc vsad} fitted position (J2000)}										 		& \multicolumn{1}{c}{2dF-radio }                & \multicolumn{1}{c}{$S_{peak}$}   & \multicolumn{1}{c}{$S_{total}$}  & \multicolumn{2}{c}{Flux uncertainty}                   & \multicolumn{1}{c}{$\sigma_n^\clubsuit$}  & \multicolumn{1}{c}{$\sigma_s^\diamondsuit$} \\
\multicolumn{1}{c}{$\alpha$} & \multicolumn{1}{c}{$\delta$}  & \multicolumn{1}{c}{}          & \multicolumn{1}{c}{}   & \multicolumn{1}{c}{}                        & \multicolumn{1}{c}{}	 & \multicolumn{1}{c}{}              & \multicolumn{1}{c}{}					& \multicolumn{1}{c}{$\alpha$} & \multicolumn{1}{c}{$\delta$} & \multicolumn{1}{c}{$\Delta^\heartsuit$}	& \multicolumn{1}{c}{sep ($^{\prime\prime})$}   & \multicolumn{1}{c}{(mJy)} 	    & \multicolumn{1}{c}{(mJy)}  	   & \multicolumn{1}{c}{(\%)} & \multicolumn{1}{c}{(mJy)}   & \multicolumn{1}{c}{(mJy)} 			     & \multicolumn{1}{c}{}					    \\
\hline
$23^{\rm h}33^{\rm m}34.07^{\rm s}$ & $-27^\circ44^\prime41.1^{\prime\prime}$ & TGS192Z064 & 0.0547    & 17.27  & 4   & a+e    & 4  & $ 			  - 				  $ & $ 			  - 					  $ &  -    & - 		 & -		  & -		   & -  		& - 		& 0.8759	   & -  		\\
$23^{\rm h}33^{\rm m}36.31^{\rm s}$ & $-28^\circ16^\prime02.2^{\prime\prime}$ & TGS270Z051 & 0.1042    & 18.02  & 4   & abs    & 4  & $ 			  - 				  $ & $ 			  - 					  $ &  -    & - 		 & -		  & -		   & -  		& - 		& 0.8026	   & -  		\\
$23^{\rm h}33^{\rm m}37.29^{\rm s}$ & $-27^\circ43^\prime20.1^{\prime\prime}$ & TGS192Z058 & 0.0546    & 16.82  & 5   & a+e?   & 4  & $ 			  - 				  $ & $ 			  - 					  $ &  -    & - 		 & -		  & -		   & -  		& - 		& 0.7719	   & -  		\\
$23^{\rm h}33^{\rm m}37.57^{\rm s}$ & $-28^\circ00^\prime42.4^{\prime\prime}$ & TGS270Z050 & 0.1055    & 17.20  & 4   & a+e    & 4  & $ 			  - 				  $ & $ 			  - 					  $ &  -    & - 		 & -		  & -		   & -  		& - 		& 0.7551	   & -  		\\
$23^{\rm h}33^{\rm m}39.31^{\rm s}$ & $-28^\circ10^\prime03.3^{\prime\prime}$ & TGS192Z060 & 0.1073    & 18.61  & 4   & SF     & 4  & $ 			  - 				  $ & $ 			  - 					  $ &  -    & - 		 & -		  & -		   & -  		& - 		& 0.7094	   & -  		\\
$23^{\rm h}33^{\rm m}39.96^{\rm s}$ & $-27^\circ51^\prime06.5^{\prime\prime}$ & TGS192Z059 & 0.0661    & 18.94  & 4   & SF?    & 4  & $ 			  - 				  $ & $ 			  - 					  $ &  -    & - 		 & -		  & -		   & -  		& - 		& 0.7028	   & -  		\\
$23^{\rm h}33^{\rm m}42.84^{\rm s}$ & $-28^\circ16^\prime04.7^{\prime\prime}$ & TGS192Z061 & 0.1042    & 18.89  & 4   & abs?   & 4  & $ 			  - 				  $ & $ 			  - 					  $ &  -    & - 		 & -		  & -		   & -  		& - 		& 0.6309	   & -  		\\
$23^{\rm h}33^{\rm m}42.94^{\rm s}$ & $-28^\circ15^\prime48.2^{\prime\prime}$ & TGS270Z048 & 0.1042    & 18.45  & 4   & abs?   & 4  & $ 			  - 				  $ & $ 			  - 					  $ &  -    & - 		 & -		  & -		   & -  		& - 		& 0.6260	   & -  		\\
$23^{\rm h}33^{\rm m}43.63^{\rm s}$ & $-28^\circ21^\prime08.8^{\prime\prime}$ & TGS192Z062 & 0.1054    & 18.73  & 4   & a+e    & 4  & $ 			  - 				  $ & $ 			  - 					  $ &  -    & - 		 & -		  & -		   & -  		& - 		& 0.6702	   & -  		\\
$23^{\rm h}33^{\rm m}48.94^{\rm s}$ & $-27^\circ25^\prime25.1^{\prime\prime}$ & TGS192Z055 & 0.1504    & 19.31  & 4   & em     & 4  & $ 			  - 				  $ & $ 			  - 					  $ &  -    & - 		 & -		  & -		   & -  		& - 		& 0.5655	   & -  		\\
$23^{\rm h}33^{\rm m}51.17^{\rm s}$ & $-27^\circ53^\prime20.8^{\prime\prime}$ & TGS192Z056 & 0.2159    & 19.38  & 1   & ???    & 4  & $ 			  - 				  $ & $ 			  - 					  $ &  -    & - 		 & -		  & -		   & -  		& - 		& 0.4698	   & -  		\\
$23^{\rm h}33^{\rm m}52.60^{\rm s}$ & $-28^\circ25^\prime05.2^{\prime\prime}$ & TGS270Z042 & 0.1064    & 18.82  & 4   & a+e    & 4  & $ 			  - 				  $ & $ 			  - 					  $ &  -    & - 		 & -		  & -		   & -  		& - 		& 0.5768	   & -  		\\
$23^{\rm h}33^{\rm m}52.97^{\rm s}$ & $-27^\circ39^\prime41.6^{\prime\prime}$ & TGS192Z054 & 0.1049    & 17.46  & 5   & abs    & 4  & $ 			  - 				  $ & $ 			  - 					  $ &  -    & - 		 & -		  & -		   & -  		& - 		& 0.4482	   & -  		\\
$23^{\rm h}33^{\rm m}54.13^{\rm s}$ & $-28^\circ20^\prime52.0^{\prime\prime}$ & TGS192Z057 & 0.2162    & 18.69  & 1   & ???    & 4  & $ 			  - 				  $ & $ 			  - 					  $ &  -    & - 		 & -		  & -		   & -  		& - 		& 0.4767	   & -  		\\
$23^{\rm h}33^{\rm m}58.07^{\rm s}$ & $-27^\circ42^\prime07.0^{\prime\prime}$ & TGS192Z052 & 0.2053    & 19.35  & 1   & ???    & 4  & $ 			  - 				  $ & $ 			  - 					  $ &  -    & - 		 & -		  & -		   & -  		& - 		& 0.3877	   & -  		\\
$23^{\rm h}33^{\rm m}58.30^{\rm s}$ & $-27^\circ23^\prime23.0^{\prime\prime}$ & TGS192Z051 & 0.1012    & 18.14  & 4   & abs    & 4  & $ 			  - 				  $ & $ 			  - 					  $ &  -    & - 		 & -		  & -		   & -  		& - 		& 0.4536	   & -  		\\
$23^{\rm h}34^{\rm m}01.04^{\rm s}$ & $-27^\circ42^\prime59.5^{\prime\prime}$ & TGS192Z053 & 0.2157    & 18.24  & 1   & ???    & 5  & $ 			  - 				  $ & $ 			  - 					  $ &  -    & - 		 & 9.500	  & 54.9	   & 8  		& 4.640 	& 0.3581	   & 27 		\\
$23^{\rm h}34^{\rm m}01.28^{\rm s}$ & $-28^\circ25^\prime53.6^{\prime\prime}$ & TGS270Z037 & 0.1537    & 18.94  & 4   & abs    & 4  & $ 			  - 				  $ & $ 			  - 					  $ &  -    & - 		 & -		  & -		   & -  		& - 		& 0.4709	   & -  		\\
$23^{\rm h}34^{\rm m}01.40^{\rm s}$ & $-27^\circ17^\prime47.3^{\prime\prime}$ & TGS192Z049 & 0.1057    & 18.85  & 4   & em     & 4  & $ 			  - 				  $ & $ 			  - 					  $ &  -    & - 		 & -		  & -		   & -  		& - 		& 0.5139	   & -  		\\
$23^{\rm h}34^{\rm m}01.43^{\rm s}$ & $-28^\circ33^\prime08.7^{\prime\prime}$ & TGS270Z039 & 0.1049    & 18.48  & 5   & abs    & 4  & $ 			  - 				  $ & $ 			  - 					  $ &  -    & - 		 & -		  & -		   & -  		& - 		& 0.7917	   & -  		\\
$23^{\rm h}34^{\rm m}01.90^{\rm s}$ & $-28^\circ27^\prime39.9^{\prime\prime}$ & TGS270Z038 & 0.1045    & 19.05  & 5   & SF     & 4  & $ 			  - 				  $ & $ 			  - 					  $ &  -    & - 		 & -		  & -		   & -  		& - 		& 0.5096	   & -  		\\
$23^{\rm h}34^{\rm m}02.61^{\rm s}$ & $-27^\circ36^\prime31.1^{\prime\prime}$ & TGS192Z050 & 0.1522    & 19.10  & 4   & abs    & 4  & $ 			  - 				  $ & $ 			  - 					  $ &  -    & - 		 & -		  & -		   & -  		& - 		& 0.3436	   & -  		\\
$23^{\rm h}34^{\rm m}03.35^{\rm s}$ & $-28^\circ15^\prime46.3^{\prime\prime}$ & TGS270Z036 & 0.2764    & 19.38  & 4   & abs?   & 3  & $ 			  - 				  $ & $ 			  - 					  $ &  -    & - 		 & 0.459	  & -		   & 157		& 0.721 	& 0.3447	   & 1  		\\
$23^{\rm h}34^{\rm m}04.46^{\rm s}$ & $-28^\circ23^\prime19.5^{\prime\prime}$ & TGS270Z032 & 0.1500    & 18.95  & 4   & ???    & 4  & $ 			  - 				  $ & $ 			  - 					  $ &  -    & - 		 & -		  & -		   & -  		& - 		& 0.3941	   & -  		\\
$23^{\rm h}34^{\rm m}04.77^{\rm s}$ & $-28^\circ24^\prime55.8^{\prime\prime}$ & TGS270Z033 & 0.1222    & 18.78  & 5   & abs    & 4  & $ 			  - 				  $ & $ 			  - 					  $ &  -    & - 		 & -		  & -		   & -  		& - 		& 0.4142	   & -  		\\
$23^{\rm h}34^{\rm m}07.88^{\rm s}$ & $-28^\circ35^\prime41.6^{\prime\prime}$ & TGS271Z031 & 0.1539    & 18.72  & 4   & em?    & 4  & $ 			  - 				  $ & $ 			  - 					  $ &  -    & - 		 & -		  & -		   & -  		& - 		& 0.8620	   & -  		\\
$23^{\rm h}34^{\rm m}08.24^{\rm s}$ & $-27^\circ22^\prime26.0^{\prime\prime}$ & TGS192Z047 & 0.1002    & 19.36  & 4   & SF?    & 4  & $ 			  - 				  $ & $ 			  - 					  $ &  -    & - 		 & -		  & -		   & -  		& - 		& 0.3667	   & -  		\\
$23^{\rm h}34^{\rm m}09.80^{\rm s}$ & $-27^\circ18^\prime32.6^{\prime\prime}$ & TGS192Z045 & 0.1006    & 19.17  & 4   & SF     & 4  & $ 			  - 				  $ & $ 			  - 					  $ &  -    & - 		 & -		  & -		   & -  		& - 		& 0.4066	   & -  		\\
$23^{\rm h}34^{\rm m}09.89^{\rm s}$ & $-28^\circ30^\prime52.3^{\prime\prime}$ & TGS270Z031 & 0.0863    & 18.74  & 5   & abs    & 3  & $ 			  - 				  $ & $ 			  - 					  $ &  -    & - 		 & 0.565	  & -		   & 166		& 0.935 	& 0.5142	   & 1  		\\
$23^{\rm h}34^{\rm m}12.46^{\rm s}$ & $-27^\circ10^\prime18.8^{\prime\prime}$ & TGS192Z185 & 0.1203    & 19.08  & 4   & a+e?   & 4  & $ 			  - 				  $ & $ 			  - 					  $ &  -    & - 		 & -		  & -		   & -  		& - 		& 0.6626	   & -  		\\
$23^{\rm h}34^{\rm m}15.03^{\rm s}$ & $-28^\circ26^\prime34.0^{\prime\prime}$ & TGS270Z030 & 0.1219    & 18.88  & 5   & SF?    & 4  & $ 			  - 				  $ & $ 			  - 					  $ &  -    & - 		 & -		  & -		   & -  		& - 		& 0.3558	   & -  		\\
$23^{\rm h}34^{\rm m}15.22^{\rm s}$ & $-27^\circ28^\prime37.1^{\prime\prime}$ & TGS192Z046 & 0.1013    & 17.92  & 4   & a+e    & 4  & $ 			  - 				  $ & $ 			  - 					  $ &  -    & - 		 & -		  & -		   & -  		& - 		& 0.2770	   & -  		\\
$23^{\rm h}34^{\rm m}15.74^{\rm s}$ & $-27^\circ27^\prime12.7^{\prime\prime}$ & TGS192Z042 & 0.1947    & 19.28  & 1   & ???    & 4  & $ 			  - 				  $ & $ 			  - 					  $ &  -    & - 		 & -		  & -		   & -  		& - 		& 0.2810	   & -  		\\
$23^{\rm h}34^{\rm m}17.09^{\rm s}$ & $-27^\circ53^\prime32.6^{\prime\prime}$ & TGS192Z044 & 0.1598    & 19.00  & 4   & ???    & 4  & $ 			  - 				  $ & $ 			  - 					  $ &  -    & - 		 & -		  & -		   & -  		& - 		& 0.2465	   & -  		\\
$23^{\rm h}34^{\rm m}20.88^{\rm s}$ & $-28^\circ32^\prime31.2^{\prime\prime}$ & TGS270Z029 & 0.1051    & 18.77  & 4   & SF     & 4  & $ 			  - 				  $ & $ 			  - 					  $ &  -    & - 		 & -		  & -		   & -  		& - 		& 0.4563	   & -  		\\
$23^{\rm h}34^{\rm m}20.91^{\rm s}$ & $-28^\circ09^\prime43.1^{\prime\prime}$ & TGS270Z028 & 0.2159    & 19.03  & 1   & ???    & 4  & $ 			  - 				  $ & $ 			  - 					  $ &  -    & - 		 & -		  & -		   & -  		& - 		& 0.2301	   & -  		\\
$23^{\rm h}34^{\rm m}23.46^{\rm s}$ & $-27^\circ15^\prime57.0^{\prime\prime}$ & TGS192Z040 & 0.0882    & 18.06  & 4   & abs    & 4  & $ 			  - 				  $ & $ 			  - 					  $ &  -    & - 		 & -		  & -		   & -  		& - 		& 0.3524	   & -  		\\
$23^{\rm h}34^{\rm m}23.60^{\rm s}$ & $-27^\circ38^\prime06.6^{\prime\prime}$ & TGS192Z041 & 0.1053    & 19.27  & 4   & abs?   & 4  & $ 			  - 				  $ & $ 			  - 					  $ &  -    & - 		 & -		  & -		   & -  		& - 		& 0.2183	   & -  		\\
$23^{\rm h}34^{\rm m}23.75^{\rm s}$ & $-27^\circ58^\prime10.2^{\prime\prime}$ & TGS270Z026 & 0.2120    & 19.02  & 4   & a+e?   & 4  & $ 			  - 				  $ & $ 			  - 					  $ &  -    & - 		 & -		  & -		   & -  		& - 		& 0.2149	   & -  		\\
$23^{\rm h}34^{\rm m}26.96^{\rm s}$ & $-28^\circ05^\prime25.7^{\prime\prime}$ & TGS270Z027 & 0.2159    & 17.84  & 1   & ???    & 4  & $ 			  - 				  $ & $ 			  - 					  $ &  -    & - 		 & -		  & -		   & -  		& - 		& 0.2030	   & -  		\\
$23^{\rm h}34^{\rm m}32.20^{\rm s}$ & $-27^\circ55^\prime10.0^{\prime\prime}$ & TGS192Z037 & -	       & 18.64  & 1   & ???    & 4  & $ 			  - 				  $ & $ 			  - 					  $ &  -    & - 		 & -		  & -		   & -  		& - 		& 0.1836	   & -  		\\
$23^{\rm h}34^{\rm m}32.30^{\rm s}$ & $-27^\circ31^\prime13.1^{\prime\prime}$ & TGS192Z036 & 0.0860    & 19.00  & 4   & abs?   & 4  & $ 			  - 				  $ & $ 			  - 					  $ &  -    & - 		 & -		  & -		   & -  		& - 		& 0.1957	   & -  		\\
$23^{\rm h}34^{\rm m}33.59^{\rm s}$ & $-28^\circ40^\prime19.7^{\prime\prime}$ & TGS270Z025 & 0.1041    & 17.82  & 5   & abs    & 4  & $ 			  - 				  $ & $ 			  - 					  $ &  -    & - 		 & -		  & -		   & -  		& - 		& 0.8460	   & -  		\\
$23^{\rm h}34^{\rm m}35.18^{\rm s}$ & $-27^\circ59^\prime43.0^{\prime\prime}$ & TGS192Z038 & 0.0866    & 17.25  & 4   & abs    & 4  & $ 			  - 				  $ & $ 			  - 					  $ &  -    & - 		 & -		  & -		   & -  		& - 		& 0.1743	   & -  		\\
$23^{\rm h}34^{\rm m}38.77^{\rm s}$ & $-28^\circ17^\prime17.1^{\prime\prime}$ & TGS270Z022 & 0.1836    & 19.20  & 5   & abs    & 4  & $ 			  - 				  $ & $ 			  - 					  $ &  -    & - 		 & -		  & -		   & -  		& - 		& 0.1797	   & -  		\\
$23^{\rm h}34^{\rm m}39.41^{\rm s}$ & $-28^\circ01^\prime37.3^{\prime\prime}$ & TGS192Z034 & 0.2121    & 19.34  & 4   & SF?    & 4  & $ 			  - 				  $ & $ 			  - 					  $ &  -    & - 		 & -		  & -		   & -  		& - 		& 0.1619	   & -  		\\
$23^{\rm h}34^{\rm m}40.05^{\rm s}$ & $-27^\circ49^\prime47.9^{\prime\prime}$ & TGS192Z033 & 0.0278    & 16.87  & 4   & SF?    & 4  & $ 			  - 				  $ & $ 			  - 					  $ &  -    & - 		 & -		  & -		   & -  		& - 		& 0.1600	   & -  		\\
$23^{\rm h}34^{\rm m}41.24^{\rm s}$ & $-28^\circ31^\prime37.1^{\prime\prime}$ & TGS270Z023 & 0.0370    & 19.07  & 4   & ???    & 4  & $ 			  - 				  $ & $ 			  - 					  $ &  -    & - 		 & -		  & -		   & -  		& - 		& 0.3015	   & -  		\\
$23^{\rm h}34^{\rm m}45.31^{\rm s}$ & $-28^\circ34^\prime52.7^{\prime\prime}$ & TGS271Z028 & 0.1054    & 18.12  & 4   & SF?    & 3  & $ 			  - 				  $ & $ 			  - 					  $ &  -    & - 		 & 0.446	  & -		   & 161		& 0.719 	& 0.3641	   & 1  		\\
$23^{\rm h}34^{\rm m}45.98^{\rm s}$ & $-27^\circ51^\prime34.0^{\prime\prime}$ & TGS192Z032 & 0.2161    & 19.31  & 1   & ???    & 3  & $ 			  - 				  $ & $ 			  - 					  $ &  -    & - 		 & 0.166	  & -		   & 164		& 0.272 	& 0.1444	   & 1  		\\
$23^{\rm h}34^{\rm m}49.29^{\rm s}$ & $-28^\circ13^\prime26.9^{\prime\prime}$ & TGS192Z031 & 0.2159    & 18.21  & 1   & ???    & 4  & $ 			  - 				  $ & $ 			  - 					  $ &  -    & - 		 & -		  & -		   & -  		& - 		& 0.1470	   & -  		\\
$23^{\rm h}34^{\rm m}50.81^{\rm s}$ & $-28^\circ38^\prime32.4^{\prime\prime}$ & TGS270Z019 & 0.1040    & 18.93  & 3   & em?    & 2  & $ 			  - 				  $ & $ 			  - 					  $ &  -    & - 		 & 2.532	  & -		   & 52 		& 1.328 	& 0.4878	   & 5  		\\
$23^{\rm h}34^{\rm m}51.62^{\rm s}$ & $-28^\circ41^\prime20.6^{\prime\prime}$ & TGS270Z020 & 0.1503    & 17.94  & 4   & abs    & 4  & $ 			  - 				  $ & $ 			  - 					  $ &  -    & - 		 & -		  & -		   & -  		& - 		& 0.7119	   & -  		\\
$23^{\rm h}34^{\rm m}55.19^{\rm s}$ & $-27^\circ01^\prime47.2^{\prime\prime}$ & TGS192Z166 & 0.1848    & 19.30  & 3   & ???    & 4  & $ 			  - 				  $ & $ 			  - 					  $ &  -    & - 		 & -		  & -		   & -  		& - 		& 0.8262	   & -  		\\
$23^{\rm h}34^{\rm m}58.39^{\rm s}$ & $-28^\circ38^\prime18.7^{\prime\prime}$ & TGS270Z018 & 0.1497    & 18.65  & 4   & abs    & 4  & $ 			  - 				  $ & $ 			  - 					  $ &  -    & - 		 & -		  & -		   & -  		& - 		& 0.4269	   & -  		\\
$23^{\rm h}35^{\rm m}01.28^{\rm s}$ & $-28^\circ43^\prime13.5^{\prime\prime}$ & TGS271Z026 & 0.1494    & 19.36  & 4   & abs    & 4  & $ 			  - 				  $ & $ 			  - 					  $ &  -    & - 		 & -		  & -		   & -  		& - 		& 0.8413	   & -  		\\
$23^{\rm h}35^{\rm m}01.81^{\rm s}$ & $-28^\circ43^\prime02.3^{\prime\prime}$ & TGS271Z025 & 0.1500    & 18.44  & 5   & abs    & 4  & $ 			  - 				  $ & $ 			  - 					  $ &  -    & - 		 & -		  & -		   & -  		& - 		& 0.8085	   & -  		\\
$23^{\rm h}35^{\rm m}03.43^{\rm s}$ & $-27^\circ18^\prime50.2^{\prime\prime}$ & TGS193Z018 & 0.1260    & 19.34  & 4   & SF     & 3  & $ 			  - 				  $ & $ 			  - 					  $ &  -    & - 		 & 0.199	  & -		   & 164		& 0.327 	& 0.1730	   & 1  		\\
$23^{\rm h}35^{\rm m}04.05^{\rm s}$ & $-27^\circ17^\prime16.4^{\prime\prime}$ & TGS192Z029 & 0.1273    & 19.24  & 4   & abs    & 4  & $ 			  - 				  $ & $ 			  - 					  $ &  -    & - 		 & -		  & -		   & -  		& - 		& 0.1850	   & -  		\\
$23^{\rm h}35^{\rm m}04.65^{\rm s}$ & $-27^\circ25^\prime15.5^{\prime\prime}$ & TGS192Z030 & 0.1246    & 17.76  & 1   & ???    & 4  & $ 			  - 				  $ & $ 			  - 					  $ &  -    & - 		 & -		  & -		   & -  		& - 		& 0.1350	   & -  		\\
$23^{\rm h}35^{\rm m}06.53^{\rm s}$ & $-27^\circ14^\prime47.6^{\prime\prime}$ & TGS192Z028 & 0.0510    & 17.76  & 5   & SF     & 4  & $ 			  - 				  $ & $ 			  - 					  $ &  -    & - 		 & -		  & -		   & -  		& - 		& 0.2055	   & -  		\\
\hline\\
\end{tabular}
\begin{tabular}{l}
$^\spadesuit$ The fitting method used to derive the radio flux or the upper-limit, (1) {\sc vsad} $4\sigma$ match, see $\S3.2$; 
              (2) pixel sum, see $\S3.3$; (3) non-identification upper-limit, see $\S3.4$;\\
              \hspace{3mm}(4) {\sc imfit} upper-limit, see $\S3.5$; (5) local noise upper-limit, see $\S3.5$\\
$^\heartsuit$   The estimated {\sc vsad} fitted positional uncertainty in arcsec, see Appendix I.\\
$^\clubsuit$    The estimated local noise, see $\S3.1.2$.\\
$^\diamondsuit$ The signal-to-noise ratio ($S_{peak}/\sigma_s$).\\
\end{tabular}\\\\\\
\end{sidewaystable}
\end{onecolumn}

\end{document}